%% file: witnesspaper6.tex
\documentclass[
notitlepage,
letterpaper,
superscriptaddress,
twocolumn,
nofootinbib,aps,pra,
10pt]{revtex4-1}

\pdfoutput=1

\usepackage{amssymb,amsmath,amsfonts,amsthm,amscd}
\usepackage{dsfont}
\usepackage{mathtools}
\usepackage{graphicx}
\usepackage{enumerate}
\usepackage{natbib}
\usepackage[
colorlinks=true, 
linkcolor=black, 
citecolor=blue, 
urlcolor=blue, 
hyperindex,
breaklinks
]{hyperref}
\usepackage{color}

\usepackage[utf8]{inputenc}
\usepackage[T1]{fontenc}
\usepackage{lmodern}

\usepackage{blindtext}

\usepackage{tikz}
\usetikzlibrary{arrows}

\theoremstyle{plain}
\newtheorem{theorem}{Theorem}

\newtheorem{proposition}[theorem]{Proposition}

\theoremstyle{definition}
\newtheorem{definition}{Definition}
\newcommand{\bra}[1]{\mathinner{\langle{#1}\rvert}}
\newcommand{\ket}[1]{\mathinner{\lvert{#1}\rangle}}
\newcommand{\ketbra}[2]{\lvert{#1}\rangle\langle{#2}\rvert}

\DeclareMathOperator{\Tr}{Tr}

\def\CC{\mathbb{C}}

\def\calC{\mathcal{C}}

\def\calE{\mathcal{E}}

\def\calI{\mathcal{I}}

\def\calN{\mathcal{N}}

\def\calS{\mathcal{S}}
\def\calT{\mathcal{T}}

\providecommand{\abs}[1]{\left\lvert#1\right\rvert}
\providecommand{\norm}[1]{\left\lVert#1\right\rVert}

\def\Gammaminus{{\Gamma_{\!-}}}
\def\Gammaplus{{\Gamma_{\!+}}}

\begin{document}

\title{Computable entanglement conversion witness that is better than the negativity}
\date{\today}
\author{Mark W.\ Girard}
\email{mwgirard (at) ucalgary.ca}
\author{Gilad Gour}
\affiliation{Institute for Quantum Science and Technology, University of Calgary}
\affiliation{Department of Mathematics and Statistics, University of Calgary, 2500 University Dr NW, Calgary, AB T2N 1N4, Canada}

\begin{abstract}
The primary goal of entanglement theory is to determine convertibility conditions for two quantum states. Up until now, this has always been done with the use of entanglement monotones. With the exception of the negativity, such quantities tend to be rather uncomputable. We instead promote the idea of conversion witnesses in this paper. A conversion witness is a function on pairs of states and whose value determines whether a state can be converted into another. We construct a conversion witness that can be efficiently computed for arbitrary states in systems of any size. This conversion witness is always better than the negativity at detecting when two entangled states are not interconvertible. Furthermore, when considering states of two-qubit systems, this new conversion witness is sometimes better than the entanglement of formation. This shows that the study of conversion witness is in fact useful, and may have applications in resource theories beyond that of entanglement.

\end{abstract}

\maketitle


\section{Introduction}

Entanglement is a necessary ingredient for many quantum information processing tasks, including the teleportation a quantum states~\cite{Bennett1993}, superdense coding~\cite{Bennett1992}, and numerous uses in quantum cryptography protocols~\cite{Bennett1984}. Two  principal features of entanglement are that it cannot be created among distant parties when there is none to begin with, and that it is depleted in the implementation of such protocols. Not all entanglement is created equal. Some entangled states may be more useful for certain applications than other entangled states. It is therefore of great interest to develop a detailed understanding of the properties of entanglement in terms of its nature as a \emph{resource}~\cite{Plenio2007}.

In the paradigmatic setting for the study of entanglement, distant parties jointly share a state of a composite quantum system. Procedures that can be performed in such a setting are limited to those that can be implemented through local operations (LO) on the subsystems and exchange of classical communication (CC) between the parties. The primary goal of entanglement theory is to fully understand the structure of the entangled states that is induced by this restriction to LOCC operations. Given two resource states $\rho$ and $\sigma$ of a joint quantum system, the fundamental question that we want to answer is the following: Can we obtain $\sigma$ from $\rho$ using only LOCC? The possible transformations of resources establishes a partial order on the set of all possible states. Fully characterizing the structure of this partial order is of chief interest, since it will allow us to determine which states will be useful for a given task.

Entanglement is typically characterized via entanglement monotones. These are functions that quantify the resourcefulness of a state, and whose value does not increase through LOCC transformations. However, it is known to be a difficult problem to determine if a given state is entangled or not~\cite{Gurvits2003}, so most monotones are unfortunately difficult to compute in general~\cite{Huang2014}. Hence, finding entanglement monotones that can be efficiently computed, yet still yield useful information about the structure of entanglement, is an essential part of the study of entanglement theory. The best (and perhaps the only) known such monotone is the \emph{negativity}~\cite{Eisert1999}. 

Individual monotones, however, can only provide a limited amount of information about which state transformations are possible. When monotones fail to discern whether a particular transformation may be achieved, we must resort to other methods to help elucidate the partial order structure of entangled states. 

The main motivation for this paper is to illustrate the importance of the concept of \emph{conversion witnesses} (first introduced in~\cite{Gour2013c}). These are functions that can detect whether or not a particular conversion is possible, without resorting to monotones to quantify the entanglement in each state.  In fact, entanglement monotones comprise a special case of a conversion witness, since a state $\rho$ cannot be converted into $\sigma$ if $\rho$ is less entangled than $\sigma$ with respect to any monotone. Most entanglement monotones, such as the entanglement of formation~\cite{Bennett1996}, cannot be computed for states of systems of arbitrary size, so it is important to develop methods of determining convertibility of states that can be determined efficiently by a computer. 

Conversion witnesses are not perfect, however, and do not in general detect every possible transformation. A single witness typically provides either a necessary or a sufficient condition for a conversion of resources to be possible, but not both. As with monotones, a witness is only useful in practice if it can be computed efficiently for any two states in consideration.  

To demonstrate that witnesses truly can be more useful than monotones, we construct a computable conversion witness for the theory of bipartite entanglement that is better than the negativity at detecting the existence of state transformations. To show that this witness does indeed improve upon the negativity, we present examples of pairs of states that have the same negativity, yet our witness detects that one cannot be converted into the other. The fact that they are not interconvertible cannot be inferred from the negativity alone, since the states have the same level of resourcefulness with respect to the negativity. In particular, we show that an entangled pure state $\rho$ of two qubits with negativity $N(\rho)<\frac{1}{3}$ cannot be converted to an entangled Werner state $\sigma$ of any dimension with the same negativity as $\rho$. We also show that our conversion witness is better than the entanglement of formation at detecting interconvertibility of two-qubit states in certain cases, but not always.

The rest of this article is structured as follows. Section \ref{sec:prelim} begins with a review of entanglement monotones to set the stage for the introduction of conversion witnesses, followed by a brief summary of positive operators and positivity under partial transposition. The notion of conversion witnesses is introduced and their rich structure is examined in section \ref{sec:witnesses}.  Our demonstrative example of a computable entanglement conversion witness is constructed in section \ref{sec:newwitness}. Construction of this witness, which is based on the negativity, is followed by a proof that this witness is indeed an improvement over the negativity. The effectiveness of our new witness is compared against the entanglement of formation and the negativity. Analysis of this new witness is concluded with a few remarks about how further witnesses might be constructed. Section \ref{sec:conclusion} concludes with a discussion on how conversion witnesses should play an important role in the study of all resource theories beyond that of entanglement.


\section{Preliminaries}
\label{sec:prelim}
In this section, the definition of an entanglement monotone is presented in order to introduce the idea of entanglement conversion witnesses. The importance of studying operations that are positive under partial transposition (PPT) for entanglement theory is discussed, and a few important facts of positive operators are reviewed.

\subsection{Entanglement monotones}

One of the main goals of entanglement theory is to understand the structure that is induced by the restriction to LOCC operations. Essentially, given any two states $\rho$ and $\sigma$ of a bipartite system, we want to be able to answer the question: can $\rho$ be converted to $\sigma$ via LOCC operations? If such a transition is possible, this is denoted as $\rho\mapsto\sigma$. Arbitrary compositions of LOCC operations are again LOCC operations. That is, if $\rho\mapsto\sigma$ and $\sigma\mapsto\tau$, then also $\rho\mapsto\tau$. Moreover, $\rho\mapsto\rho$ for any state $\rho$ by simply doing nothing. This induces a partial order on the set of states. 

Quantifying the entanglement in states is the standard method for characterizing this partial order structure in entanglement theory. If it is possible to convert a state $\rho$ into another state $\sigma$, then $\rho$ is at least as useful for any task that requires $\sigma$. Hence~$\rho$ must be at least as entangled as~$\sigma$ under any measure of entanglement.  Finding useful entanglement measures, or \emph{entanglement monotones}, is important for describing which state transformations may or may not be possible under LOCC. 

\begin{definition}
 An \emph{entanglement monotone} is a real-valued function $f$ on quantum states of bipartite systems that does not increase under LOCC operations. That is, the function $f$ is a monotone if $f(\calE(\rho))\leq f(\rho)$ for all states~$\rho$ and LOCC operations $\calE\in\mathcal{C}$.
\end{definition}

Equivalently, a monotone is a function such that $\rho\mapsto\sigma$ implies $f(\rho)\geq f(\sigma)$. If $\rho$ and $\sigma$ are states such that $f(\rho)<f(\sigma)$, then it is clear that $\rho\not\mapsto\sigma$. The entanglement monotone $f$ is said to \emph{detect} this non-convertibility. However, a single monotone does not typically supply enough information to determine if an arbitrary pair of quantum states is convertible or not.  Indeed, even if~$\rho$ is more entangled than~$\sigma$ under some monotone, it may still be the case that~$\rho$ cannot be converted into~$\sigma$.

A family of monotones $f_i$ indexed by $i\in\calI$ is said to be \emph{complete} if $\rho\mapsto\sigma$ if and only if $f_i(\rho)\geq f_i(\sigma)$ for all $i\in\calI$. A complete family of monotones can always be trivially defined. For each state $\tau$, define the function
\[
 f_\tau(\rho) = \left\{\begin{array}{ll}
                        1, & \rho\mapsto\tau\\
                        0, & \rho\not\mapsto\tau.
                       \end{array}\right.
\]
Although its value cannot be straightforwardly computed, $f_\tau$ is indeed a monotone, and the family $(f_\tau)_{\tau}$ naturally comprises a complete family of monotones. Even though a complete family of monotones exists, this family may not necessarily be useful. Given states $\rho$ and $\sigma$, one would need to be able to actually \emph{compute} the values $f(\rho)$ and $f(\sigma)$ for the monotone $f$ to be practical. 

When considering bipartite entanglement of only pure states, such a complete family of computable monotones is known to exist. Given two pure states $\ket{\psi}$ and $\ket{\phi}$ of two systems with dimension $d$, the question of convertibility can be cast in terms of majorization~\cite{Nielsen1999}. There exists an LOCC channel converting $\ket{\psi}$ into $\ket{\phi}$ if and only if $\ket{\psi}$ is majorized by $\ket{\phi}$, that is if
\[
 \sum_{i=1}^k \lambda^{(\psi)}_i \leq  \sum_{i=1}^k \lambda^{(\phi)}_i \qquad\text{for all }k=1,\dots,d,
\]
where $\lambda^{(\psi)}_i$ and $\lambda^{(\phi)}_i$ are the Schmidt coefficients of those states in decreasing order.
A complete family of entanglement monotones for pure states can be constructed from this crieterion. For each $l=2,\dots,d$, the function
\[
 f_l(\psi)=\sum_{i=l}^d\lambda_i^{(\psi)}
\]
is an entanglement monotone, and these monotones completely determine convertibility for pure states~\cite{Nielsen2001}.

For pure states of two-qubit systems, convertibility is precisely determined by another well-known entanglement monotone: the concurrence (or, equivalently, the entanglement of formation) \cite{Wootters1998}. The concurrence can also be computed explicitly for arbitrary two-qubit mixed states. However the concurrence no longer completely determines convertibility of arbitrary mixed states in a two-qubit system \cite{Miranowicz2004}. 

In systems larger than two qubits, the entanglement of formation can no longer be explicitly computed for arbitrary mixed states (although it can be computed for pure states and some mixed states with a high degree of symmetry \cite{Vollbrecht2001}). The negativity is the only known entanglement monotone that can be computed for arbitrary states of systems of any size, and its computation is only as difficult as computing eigenvalues of a matrix. Furthermore, a complete family of finitely many computable monotones cannot exist for determining convertibility of arbitrary mixed states for systems of any size~\cite{Gour2005a}. 

In light of the lack of computable entanglement monotones for arbitrary quantum states, other methods of easily determining whether a particular state transformation is possible are desired. This is exactly the purpose of this paper. In section \ref{sec:witnesses}, we present a computable method of determining state transformations for states of arbitrary size. Furthermore, the method we present is better than the negativity at detecting when a state may be transformed into another using LOCC.

\subsection{LOCC and PPT operations}

Although LOCC emerges as the natural class of operations for most tasks in quantum information, its mathematical structure is highly complex and difficult to characterize \cite{Cros2013} (for a precise definition of the LOCC class see \citep[sec.\ XI]{Horodecki2009}). Equally troublesome is the task of characterizing the separable states. In fact, the separability problem in arbitrary dimensions of the subsystems is known to be NP-Hard~\cite{Gurvits2003}. It is therefore desirable to consider other classes of operations that still provide interesting insights regarding entanglement. 

Perhaps one of the most elegant results in the early days of quantum information was the characterization of entangled states through partial transposition~\cite{Peres1996,Horodecki1996}. If a state $\rho$ of a bipartite system has $\rho^\Gamma\not\geq 0$, then $\rho$ must be entangled, where  $\Gamma$ indicates the partial transpose of $\rho$ with respect to one of the subsystems. The set of states that are positive under partial transposition (PPT) include the separable ones. Moreover, all PPT entangled states are bound entangled. That is, they are `useless' for generating entanglement~\cite{Horodecki1998} within the framework of LOCC. Hence, the set of PPT states is not only easy to characterize but also very useful in the study of entanglement~\cite{Matthews2008}.

In addition to the class of LOCC operations, we also consider the class $\calC_\Gamma$ of operations that are positive under partial transposition (PPT) \cite{Rains1999,Rains1999a}. This is the set of all completely positive trace-preserving maps $\calE$ such that the partially transposed map $\calE^\Gamma$, defined by
\[
 \calE^\Gamma(\rho):= \left[\calE(\rho^\Gamma)\right]^\Gamma, 
\]
is also completely positive. It is well known that all LOCC operations (and all separable operations~\cite{Gheorghiu2007}) form a subset of the PPT operations~\cite{Rains1999,Rains1999a,Rains2001}. Hence, any function of states that is a no-go conversion witness for convertibility under PPT operations is also a no-go entanglement conversion witness. With this fact, in section~\ref{sec:newwitness} we construct a computable no-go conversion witness based on the structure of PPT maps.

\subsection{Useful properties of positive operators}
We recall a few useful facts about positive operators. Denote by $H_n$ the set of $n\times n$ hermitian matrices. Let $H_{n,+}$ denote the cone of positive semi-definite hermitian matrices, and furthermore let $H_{n,+,1}$ denote the subset of those with unit trace. Hence $H_{n,+,1}$ is equivalent to the set of states of a $n$-dimensional quantum system.

Every operator $A\in H_n$ has a unique decomposition into its orthogonal positive and negative components,
\begin{equation}\label{eq:posnegparts}
 A=A_+-A_-,
\end{equation}
where $A_+$ and $A_-$ are the unique positive definite matrices satisfying \eqref{eq:posnegparts} such that $A_+A_-=A_-A_+=0$. Furthermore, for any two positive semi-definite operators $A,B\in H_{n,+}$, we have the following operator inequalities:
\begin{equation}
\label{eq:posmatrices}
 (A-B)_+\leq A \hspace{5mm}\text{and}\hspace{5mm}(A-B)_-\leq B
\end{equation}
(see Appendix \ref{app:opineqproofs} for proof). We also recall that the 1-norm of an hermitian operator $\norm{A}_1=\Tr\abs{A}$ is the sum of the absolute values of its eigenvalues, where the operator $\abs{A}=A_+ + A_-$ is the modulus of $A$. 

Finally, note that for positive operators $A,B\in H_{n,+}$, the operator inequality $A\leq B$ holds if and only if
\begin{equation}\label{eq:positiveoperatorsequivalentinequality}
 \Tr[\gamma A]\leq \Tr[\gamma B]\hspace{3mm}\text{for all }\gamma\in H_{n,+,1}.
\end{equation}
This useful characterization will be exploited in the construction of our conversion witnesses.


\section{Conversion witnesses}
\label{sec:witnesses}

In this section, we introduce the concept of entanglement conversion witnesses, a generalization of monotones. Conversion witnesses that can be easily computed may be more effective than monotones at determining the convertibility of quantum states. The rich structure of these conversion witnesses is also explored. 

In the previous section, it was noted that the extent of the usefulness of entanglement monotones in comparing quantum states is inherently limited, and thus other methods must be found. The most general technique for characterizing the convertibility of states is through \emph{conversion witnesses}\footnote{This concept was first introduced as \emph{relative monotones} in \cite{Sanders2010}.} (see \citep[sec.\ II.A.]{Gour2013c}).

\begin{definition}[Entanglement conversion witnesses]
Let~$W$ be a real-valued function on pairs of quantum states. If $W(\rho,\sigma)\geq0$ implies that $\rho\mapsto\sigma$ under LOCC, then $W$ is said to be a \emph{go witness}. If $W(\rho,\sigma)<0$ implies that $\rho\not\mapsto\sigma$, then $W$ is said to be a \emph{no-go witness}. Finally, $W$ is said to be a \emph{complete witness} if it is both a go and a no-go witness.
\end{definition}

Given a monotone $f$, we can define a no-go witness by $W_f(\rho,\sigma) = f(\rho)-f(\sigma)$. Indeed $W_f(\rho,\sigma)<0$ implies $f(\rho)<f(\sigma)$, and thus $\rho\not\mapsto\sigma$ by the monotonicity of $f$. Hence entanglement monotones can be considered as a special case of entanglement conversion witnesses.

The set of all no-go witnesses is endowed with the structure of a partially ordered set. Indeed, given two no-go witnesses $W_1$ and $W_2$, we say that $W_1\succeq W_2$ if
\begin{equation}
 W_2(\rho,\sigma)<0 \, \Longrightarrow \, W_1(\rho,\sigma)<0 \hspace{2mm}\text{ for all }\rho,\sigma.
\end{equation}
That is, $W_1\succeq W_2$ means that the witness $W_1$ tells us more information about the convertibility of states than $W_2$ does. If $W_2$ detects the inconvertibility $\rho\not\mapsto\sigma$ for some states $\rho$ and $\sigma$, this same information can already be obtained by $W_1$. But $W_1$ might be able to detect the inconvertibility of other pairs of states that $W_2$ cannot.

The partial order structure of no-go witnesses is illuminated in the following example: Given a family of no-go witnesses $(W_i)_{i\in\calI}$, we can construct a new no-go witness $W_\calI$ by minimizing over all witnesses in the family
\[
 W_\calI(\rho,\sigma):=\min_{i\in\calI}W_i(\rho,\sigma).
\]
This is indeed a witness, since $W_\calI(\rho,\sigma)<0$ implies that $W_i(\rho,\sigma)<0$ for at least one $i\in\calI$ and thus $\rho\not\mapsto\sigma$. Hence $W_\calI\succeq W_i$ and the resulting witness $W_\calI$ is an improvement over each of the sub-witnesses $W_i$. Similarly, given a family $(f_i)_{i\in\calI}$ of monotones, one can define a witness
\[
 W_\calI(\rho,\sigma):=\min_{i\in\calI}\left\{f_i(\rho)-f_i(\sigma)\right\} = \min_{i\in\calI}W_{f_i}(\rho,\sigma) .
\]
If the family $(f_i)_{i\in \calI}$ is complete, then the resulting $W_\calI$ is a complete witness. Furthermore, if $\overline{W}$ is a complete witness, then $\overline{W}\succeq W$ for any no-go witness $W$. 

\begin{figure}[t!]
 \centering
 \begin{tikzpicture}[->,auto,node distance=15mm,main node/.style={font=\sffamily}]

  \node[main node] (1) {$W_{\{1,2,3\}}$};
  \node[main node] (2) [below of=1] {$W_{\{1,3\}}$};
  \node[main node] (3) [left of=2] {$W_{\{1,2\}}$};
  \node[main node] (4) [right of=2] {$W_{\{2,3\}}$};
  \node[main node] (5) [below of=3] {$W_1$};
  \node[main node] (6) [below of=2] {$W_2$};
  \node[main node] (7) [below of=4] {$W_3$};

  \path[every node/.style={font=\sffamily\small}]
    (1) edge node [left] {} (2)
        edge node [left] {} (3)
        edge node [left] {} (4)
    (3) edge node [right] {} (5)
        edge node [right] {} (6)       
    (2) edge node [right] {} (5)
        edge node [right] {} (7)
    (4) edge node [right] {} (6)
        edge node [right] {} (7);
\end{tikzpicture}
\caption{An example hierarchy of no-go conversion witnesses in which an arrow between witnesses $X\longrightarrow Y$ denotes $X\succeq Y$. Consider three no-go witnesses $W_1$, $W_2$ and $W_3$, which may be incomparable with respect to the partial order. The witnesses $W_{\{1,2\}}$, $W_{\{1,3\}}$ and $W_{\{2,3\}}$ are obtained by minimizing over the sub-witnesses $W_1$, $W_2$ and $W_3$ respectively. At the top of the partial order is the witness $W_{\{1,2,3\}}$ obtained by minimizing over all three. }
\label{fig:hierarchy}

 \end{figure}
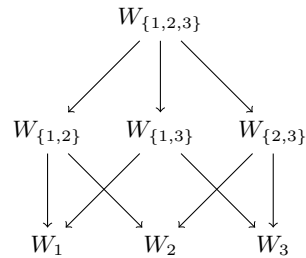

An example hierarchy of no-go conversion witnesses is depicted in FIG.\ \ref{fig:hierarchy}. Note that two no-go witnesses $W_1$ and $W_2$ may be incomparable in general. That is, it may be that both $W_1\not\succeq W_2$ and $W_2\not\succeq W_1$.   

An analogous partial order exists for \emph{go} witnesses. If $W_1$ and $W_2$ are two go witnesses and $W_1(\rho,\sigma)\geq 0 $ implies that $W_2(\rho,\sigma)\geq 0$ for all states $\rho$ and $\sigma$, then we say that $W_1\succeq W_2$. Given a family of go witnesses $(W_i)_{i\in\calI}$, a new go witness 
\[
 W_\calI(\rho,\sigma):=\max_{i\in\calI}W_i(\rho,\sigma)
\]
can be constructed such that $W_\calI\succeq W_i$ for each sub-witness. Additionally, we have $\overline{W}\succeq W$ for any complete witness $\overline{W}$ and any go witness $W$.

\section{New entanglement conversion witnesses}
\label{sec:newwitness}

In this section, we construct a computable example of a no-go conversion witness for entangled states that is based on the construction of the negativity. We first recall a few useful properties of positive operators. The proof that the negativity is a monotone under PPT operations is then reproduced, since it is needed to construct our new conversion witness. After establishing our new no-go witness, we subsequently show that this witness is, in fact, better than the negativity at detecting convertibility under LOCC. Although the conversion witness is not computable as it is initially introduced, we consider simplified versions of the witness that are computable and still supply valuable information about conversion that is not available from the negativity alone. Finally, we compare these witnesses with the negativity using the partial order structure introduced in the previous section.

\subsection{Negativity is a PPT monotone}

We recall that the \emph{negativity} \cite{Zyczkowski1998,Eisert1999,Vidal2002} of a bipartite quantum state is defined as 
\begin{equation*}
 N(\rho)= \frac{\lVert\rho^\Gamma\rVert_1-1}{2},
\end{equation*}
which is the sum of the negative eigenvalues of the partial transpose of a normalized state. The negativity is known to be a monotone under LOCC operations and hence comprises an entanglement measure that can be computed effectively. A more convenient definition for the negativity that we employ here is
\begin{equation}\label{eq:negativity}
 N(\rho):=\Tr[\rho^\Gammaminus],
\end{equation}
which does not depend on the normalization of $\rho$, and where $\rho^\Gammaminus = (\rho^\Gamma)_{-}$ represents the negative part of the partial transpose of the operator $\rho$. 

Following the proof in~\cite{Plenio2007}, we now show that the negativity defined in \eqref{eq:negativity} is a monotone under PPT operations. Let $\calE\in\calC_\Gamma$ be a PPT operation and $\rho$ be a state of a bipartite system. Note that $\calE^\Gamma(\rho^\Gamma)=[\calE(\rho^{\Gamma\Gamma})]^\Gamma = \calE(\rho)^\Gamma$ \cite{Rains1999a} and thus
\begin{align*}
 \calE(\rho)^\Gammaminus &= \left[\calE^\Gamma(\rho^\Gamma)\right]_-\\
 &= \left[\calE^\Gamma(\rho^\Gammaplus-\rho^\Gammaminus)\right]_-\\
 &=\left[\calE^\Gamma(\rho^\Gammaplus)-\calE^\Gamma(\rho^\Gammaminus)\right]_-\\
 &\leq\calE^\Gamma(\rho^\Gammaminus).
\end{align*}
The inequality in the last line follows from \eqref{eq:posmatrices}, since $\calE^\Gamma$ is a positive map and the operators $\rho^\Gammaplus=(\rho^\Gamma)_+$ and $\rho^\Gammaminus$ are both positive.  Hence, we have the operator inequality 
\begin{equation}\label{eq:pptinequality}
\calE(\rho)^\Gammaminus\leq \calE^\Gamma(\rho^\Gammaminus)
\end{equation}
 for all states $\rho$ and PPT operations $\calE$. Since $\calE$ and $\calE^\Gamma$ are trace-preserving, taking the trace of the right side of~\eqref{eq:pptinequality} yields $\Tr[\calE^\Gamma(\rho^\Gammaminus)] = \Tr[\rho^\Gammaminus] = N(\rho)$, which is just the negativity of the state $\rho$.  Taking the trace of both sides of \eqref{eq:pptinequality} yields the inequality
\begin{equation}
 N(\calE(\rho))\leq N(\rho),
\end{equation}
i.e.\ the negativity is in fact a monotone under PPT operations. Recall that the LOCC operations are a subset of PPT ones, so the negativity is a monotone under LOCC as well.

Since the negativity is a monotone under PPT operations, the convertibility condition $\rho\mapsto\sigma$ implies $N(\rho)\geq N(\sigma)$ for any states $\rho$ and $\sigma$ of a bipartite system.

\subsection{A no-go witness for PPT conversion}
We are now ready to construct no-go entanglement conversion witnesses that are based on the construction of the negativity discussed in the previous subsection. 

Let $\rho$ and $\sigma$ be arbitrary quantum states. Suppose that we want to determine if $\rho\mapsto\sigma$ under PPT operations. If $\sigma$ \emph{can} be obtained from $\rho$ via PPT operations, then $\sigma=\calE(\rho)$, and thus
\begin{equation}\label{eq:pptinequalitysigma}
\sigma^\Gammaminus \leq \calE^\Gamma\left(\rho^\Gammaminus\right)
\end{equation}
for some PPT operation $\calE\in\calC_\Gamma$. On the other hand, if $\sigma$ \emph{violates} this operator inequality for \emph{all} PPT operations, then we have the condition that
\begin{equation}\label{eq:pptinequalitysigma2}
 \sigma^\Gammaminus \not\leq \calE^\Gamma\left(\rho^\Gammaminus\right) \hspace{5mm} \text{for all } \calE\in\calC_\Gamma,
\end{equation}
and so $\sigma\neq\calE(\rho)$ for all $\calE\in\calC_\Gamma$. It follows that $\rho\not\mapsto\sigma$ under PPT operations. 

Taking the trace of both sides of \eqref{eq:pptinequalitysigma2}, this `witness' simply reduces to the statement that $N(\sigma)\not\leq N(\rho)$ implies $\rho\not\mapsto\sigma$, which already follows from the fact that the negativity is a PPT monotone. However, the operator inequalities of~\eqref{eq:pptinequalitysigma2} contain much more information that is `lost' by taking the trace and reducing it to the one-dimensional inequality $N(\sigma)\not\leq N(\rho)$. The idea that  more useful information may be extracted from this operator inequality in \eqref{eq:pptinequalitysigma2} implies that a family of no-go witnesses might be constructed from it. 

Before introducing such no-go witnesses explicitly, we present a few more useful concepts. The \emph{support function} of a subset $C\subset H_{n}$ is defined as~\cite{Weis2011}
\begin{equation}\label{eq:supportfunction}
 h_{C}(\rho) := \sup_{\gamma\in C}\Tr[\gamma\rho]
\end{equation}
for any $\rho\in H_{n}$. If $C$ is also compact, then $\sup$ in \eqref{eq:supportfunction} may be replaced with $\max$. 

We now define some sets of states that we use in the following analysis. For every $c\geq0$, define the set
\begin{equation}
\label{eq:negsets}
 \calN_c = \left\{\gamma\in H_{n,+,1}\,\middle|\, N(\gamma)\leq c \right\},
\end{equation}
which is the set of normalized states with negativity at most $c$. For $c=0$, the set $\calN_0$ is just the set of PPT states. Given a class of CPTP operations $\calC$, we can consider the `orbit' of states that are reachable from a given state~$\rho$ via operations from $\calC$. This is denoted as
\begin{equation}
 \calC(\rho) :=  \left\{\calE(\rho)\,\middle|\, \calE\in\calC \right\}.
\end{equation}

Consider the orbit of a state $\rho$ under PPT operations. All states in the orbit $\calC_\Gamma(\rho)$ must have negativity at most $N(\rho)$, since the negativity is a monotone under PPT operations $\calC_\Gamma$. This  implies the containments
\[
\calC_\Gamma(\rho) \subset \calN_{N(\rho)}
\]
for all normalized states $\rho$. Hence, the respective support functions of these sets obey the inequality
\[
h_{\calC_\Gamma(\rho)}\leq  h_{\calN_{N(\rho)}}
\]
whenever $\rho$ is a normalized state.

We now return to the task of constructing no-go witnesses for PPT conversion. For some states $\rho$ and $\sigma$, suppose that $\rho\mapsto\sigma$.  Then  $\sigma=\calE(\rho)$ and hence the operator inequality in \eqref{eq:pptinequalitysigma} holds for some $\calE\in\calC_\Gamma$. Using the property of positive operators from \eqref{eq:positiveoperatorsequivalentinequality}, the fact that the operator inequality $\sigma^\Gammaminus\leq \calE^\Gamma(\rho^\Gammaminus)$ holds for some $\calE\in\calC_\Gamma$ is equivalent to the statement that the family of inequalities
\begin{equation}\label{eq:pptinequalitysigmaerho}
\Tr[\tau\sigma^\Gammaminus] \leq \Tr[\tau\,\calE^\Gamma(\rho^\Gammaminus)] 
\end{equation}
holds for all $\tau\in H_{n,+,1}$ and some PPT operation $\calE\in\calC_\Gamma$. Note that the partial transpose operator $\calE^\Gamma$ of a PPT operation is again another PPT operation, hence
\[
 \Tr[\tau\,\calE^\Gamma(\rho^\Gammaminus)]\leq h_{\calC_\Gamma(\rho^\Gammaminus)}(\tau)
\]
for all $\tau\in H_{n,+,1}$, where $\rho^\Gammaminus$ is a non-normalized state. Therefore, if the conversion $\rho\mapsto\sigma$ is possible under PPT operations, the inequality 
\[
 \Tr[\tau\sigma^\Gammaminus]\leq h_{\calC_\Gamma(\rho^\Gammaminus)}(\tau)
\]
must hold for all $\tau$. It follows that, for each $\tau\in H_{n,+,1}$, the function 
\[
 \widehat{W}_\tau(\rho,\sigma):=h_{\calC_\Gamma(\rho^\Gammaminus)}(\tau) - \Tr[\tau\sigma^\Gammaminus]
\]
is a valid no-go conversion witness for PPT operations. That is, if $\widehat{W}_\tau(\rho,\sigma)$ is negative for some $\tau$, then it must be the case that $\rho\not\mapsto\sigma$. 

If $\rho$ itself is a PPT state, then $\rho^\Gamma\geq 0$ and thus $\rho^\Gammaminus=0$ so the negativity of $\rho$ vanishes. Hence $\sigma$ can be obtained from $\rho$ only if $\sigma$ is also a PPT state with vanishing negativity. For any interesting applications of these conversion witnesses, we may assume that the initial state $\rho$ is entangled with non-vanishing negativity and thus $\rho^\Gammaminus\neq0$.  For each non-PPT state $\rho$ we can renormalize the operator $\rho^\Gammaminus$ to define a normalized state
\begin{equation}\label{eq:definerhotilde}
 \tilde\rho = \tfrac{1}{N(\rho)}\rho^\Gammaminus,
\end{equation}
where we use the fact that $N(\rho)=\Tr[\rho^\Gammaminus]$. The inequality in \eqref{eq:pptinequalitysigmaerho} is then equivalent to 
\begin{equation*}
 \Tr[\tau \sigma^\Gammaminus] \leq N(\tilde\rho)\Tr[\tau \calE^\Gamma(\tilde\rho)].
\end{equation*}
Employing of the support function for the orbit of $\tilde\rho$ under PPT operations, we have $\Tr[\tau \calE^\Gamma(\tilde\rho)]\leq h_{\calC_\Gamma(\tilde\rho)}(\tau)$ and thus
\begin{equation}\label{eq:orbitinequality}
\Tr[\tau \sigma^\Gammaminus] \leq N(\tilde\rho)\, h_{\calC_\Gamma(\tilde\rho)}(\tau)
\end{equation}
for each $\tau\in H_{n,+,1}$ if the conversion $\rho\mapsto\sigma$ is possible.
This yields a no-go witness for each $\tau$, since $\widehat{W}_\tau(\rho,\sigma)<0$ implies $\rho\not\mapsto\sigma$, where $\widehat{W}_\tau$ is the witness defined by
\begin{equation}\label{eq:firstwitnessf}
 \widehat{W}_\tau(\rho,\sigma) = N(\tilde\rho)\, h_{\calC_\Gamma(\tilde\rho)}(\tau) - \Tr[\tau \sigma^\Gammaminus].
\end{equation}
Note that if we chose $\tau=\frac{1}{n}I$ to be the maximally mixed state, evaluating the support function simplifies to
\begin{equation}\label{eq:supportformaxmix}
 h_{\calC_\Gamma(\tilde\rho)}(\tau) = \frac{1}{n}\max_{\calE\in\calC_\Gamma} \underbrace{\Tr[\calE(\tilde\rho)]}_{=1} = \frac{1}{n}
\end{equation}
since each $\calE\in\calC_\Gamma$ is trace preserving and $\Tr\,\tilde\rho=1$. So the witness~\eqref{eq:firstwitnessf} simplifies to
\[
 \widehat{W}_{\frac{1}{n}I}(\rho,\sigma) = \frac{1}{n}\left(N(\rho)-N(\sigma)\right) = \frac{1}{n}W_N(\rho,\sigma),
\]
which is just the difference of negativities of the two states. Thus, the family of witnesses $\widehat{W}_\tau$ yields at least as much information regarding the convertibility as the negativity does. In fact,  minimizing $\widehat{W}_\tau$ over all possible $\tau\in H_{n,+,1}$ yields the witness
\[
 \widehat{W}(\rho,\sigma) : = n \min_{\tau} \widehat{W}_{\tau}(\rho,\sigma)
\]
such that $\widehat{W}(\rho,\sigma)\leq N(\rho)-N(\sigma)$ for all states $\rho$ and $\sigma$. In the syntax of the partial order of no-go witnesses, we have that $\widehat{W}\succeq \widehat{W}_\tau$ for each $\tau$, and $\widehat{W}\succeq W_N$, where $W_N$ is the witness formed from the negativity. This means that the witness $\widehat{W}$ may be able to yield more information about the convertibility of arbitrary states under PPT operations than the best-known monotone, the negativity.

However, the support function $h_{\calC_\Gamma(\tilde\rho)}(\tau)$ is very difficult to calculate in general, since it involves an optimization over all PPT operations. So this witness is not very computable in practice. In the following, we construct a more useful generalization of this witness. 

Replacing the orbit $\calC_\Gamma(\tilde\rho)$ in~\eqref{eq:orbitinequality} with the set $\calN_{N(\tilde\rho)}$ (i.e.\ the set of states whose negativity is bounded by $N(\tilde\rho)$) yields the inequality
\[
 \Tr[\tau\sigma^\Gammaminus]\leq N(\rho)h_{\calN_{N(\tilde\rho)}}(\tau).
\]
Indeed, $\calC_\Gamma(\tilde\rho)\subset \calN_{N(\tilde\rho)}$ and thus $h_{\calC_\Gamma(\tilde\rho)}(\tau)\leq h_{\calN_{N(\tilde\rho)}}(\tau)$. This inequality must hold for all $\tau$ if $\rho\mapsto\sigma$. Hence, we obtain a family of witnesses defined by
\begin{equation}\label{eq:witnesstau}
 W_{\tau}(\rho,\sigma) = N(\rho)\, h_{\calN_{N(\tilde\rho)}}(\tau) - \Tr[\tau \sigma^\Gammaminus]
\end{equation}
for each $\tau\in H_{n,+,1}$. If $W_{\tau}(\rho,\sigma)<0$ for any $\tau$, then $\rho\not\mapsto\sigma$. Note that $W_{\tau}(\rho,\sigma)\leq \widehat{W}_\tau(\rho,\sigma)$, so the witnesses $W_{\tau}$ supply less information about convertibility than the witnesses $\widehat{W}_\tau$ do. However, choosing $\tau=\frac{1}{n}I$, we see that $h_{\calN_{N(\tilde\rho)}}(\frac{1}{n}I)= 1$. Thus 
\[
nW_{\frac{1}{n}I}(\rho,\sigma) = N(\rho)-N(\sigma) = W_N(\sigma,\rho),
\]
 so the new witnesses $W_{\tau}$ still yields at least as much information as the difference of negativities. As before, minimizing over all $\tau\in H_{n,+,1}$ yields the witness
\begin{equation}\label{eq:witnessf1}
 W(\rho,\sigma):=n \min_\tau W_{\tau}(\rho,\sigma)
\end{equation}
such that $W(\rho,\sigma)\leq N(\rho)-N(\sigma)$. In particular, we have $W\succeq W_\tau$ for each $\tau$. Furthermore, the hierarchy of no-go witnesses $\widehat{W}\succeq W\succeq W_N$ holds. 

This witness is still not very useful in practice, since $h_{\calN_c}(\tau)$ is difficult to compute for arbitrary $\tau$. In the following section, we show how to compute $h_{\calN_c}(\tau)$ for certain highly symmetric states in order to construct computable versions of the witnesses $W_{\tau}$ and $W$.

\subsection{A computable no-go witness}
\label{sec:computablenogowitness}

Although the support function $h_{\calN_c}(\tau)$ cannot be determined in general for arbitrary $\tau$, it can be evaluated explicitly for certain operators $\tau$ that exhibit high degrees of symmetry, such as the Werner states and isotropic states \cite{Horodecki1999,Vollbrecht2001,Vidal2002}. Rather than performing the minimization in \eqref{eq:witnessf1} over \emph{all} states $\tau$, we can instead minimize over classes of states for which $h_{\calN_c}(\tau)$ is computable. 

Recall from \eqref{eq:supportformaxmix} that $h_{\calN_c}(\tau)$ is computable for $\tau=\frac{1}{n}I$, and that $W_{\frac{1}{n}I}(\rho,\sigma)<0$ if $N(\rho)-N(\sigma)<0$. Thus, restricting the minimization in \eqref{eq:witnessf1} to only be over states where $h_{\calN_c}$ is computable will still yield a witness that is at least as good as the difference of negativities. In this section, we perform such a minimization over a small class of states to construct a computable example of a no-go witness.

We restrict to states of a bipartite $d\times d$-system, where $H_{n,+,1}$ comprises the states of the system and $n=d^2$. The Werner states are those that are invariant under all unitaries of the form $U\otimes U$, where $U$ is any unitary on the $d$-dimensional subsystems. Furthermore, the Werner states are invariant under application of the twirling operation of the form
\[
 \calT_{U\otimes  U} (\rho) = \frac{1}{\int dU}\int U\otimes {U}\, \rho \, U^\dagger\otimes U^\dagger \, dU,
\]
where $dU$ denotes the standard Haar measure on the group of all $d\times d$ unitary matrices. Not only do we have  $\calT_{U\otimes U}(\rho)=\rho$ for any Werner state $\rho$, but applying $\calT_{U\otimes U}$ to any state always yields an Werner state. The Werner states on a $d\times d$-system form a one-dimensional family that may be parametrized by
\begin{equation*}
 p\frac{1}{d}F + \frac{1-p}{d^2} I \hspace{2mm}\text{ for }\tfrac{-1}{d-1}\leq p \leq \tfrac{1}{d+1},
\end{equation*}
where $F$ is the so-called `flip' operator on a $d\times d$-system such that $F\ket{\psi}\otimes\ket{\phi}=\ket{\phi}\otimes\ket{\psi}$ for all product vectors. Explicitly, the flip operator is defined by
\[
 F=\sum_{i,j=1}^d\ketbra{ij}{ji}.
\]
Note that this can also be given by $F=d\ketbra{\Phi}{\Phi}^\Gamma$, where $\ket{\Phi}$ is the maximally entangled state of a $d\times d$-system,
\begin{equation}\label{eq:maxentdxd}
 \ket{\Phi}=\frac{1}{\sqrt{d}}\sum_{j=1}^d \ket{j}\otimes\ket{j}.
\end{equation}
It is more convenient to parametrize the Werner states with respect to a more dimension-independent parameter as
\begin{equation}\label{eq:wernerparameter}
 \omega^d_\alpha = \frac{d\alpha-1}{d(d^2-1)}F + \frac{d-\alpha}{d(d^2-1)}I, \qquad -1\leq \alpha \leq 1.
\end{equation}
This parameter is exactly the overlap of the Werner state with the flip operator is $\alpha = \Tr[\omega_\alpha F]$. With this parametrization, the Werner states with parameter $\alpha$ are entangled exactly when $\alpha<0$.

Another family of states with a high degree of symmetry that is typically studied in bipartite entanglement consists of the \emph{isotropic states}. Similar to the Werner states, the isotropic states are invariant under all unitaries of the form $U\otimes \bar{U}$, where the $\bar{U}$ denotes the complex conjugate of $U$. The isotropic states are invariant under the application of a twirling operation of the form
\[
 \calT_{U\otimes \bar U} (\rho) = \frac{1}{\int dU}\int U\otimes \bar{U}\, \rho \, U^\dagger\otimes \bar{U}^\dagger \, dU,
\]
such that $\calT_{U\otimes\bar U}(\rho)=\rho$ for any isotropic state $\rho$. Applying $\calT_{U\otimes\bar U}$ to any state always yields an isotropic state. The isotropic states of a $d\times d$-system can be parametrized as
\begin{equation}\label{eq:eqisotropic}
 \eta^d_\beta=\beta\ketbra{\Phi}{\Phi} + \frac{1-\beta}{d^2-1}\bigl(I-\ketbra{\Phi}{\Phi}\bigr), \qquad 0\leq \beta\leq 1
\end{equation}
where the parameter $\beta$ corresponds to the overlap with the maximally entangled state $\beta = \Tr\left[\eta_\beta\ketbra{\Phi}{\Phi}\right]$.

Making use of this high degree of symmetry, the support function $h_{\calN_c}(\tau)$ can be explicitly evaluated for these families of states. Indeed, when the state $\tau=\omega_\alpha^d$ is Werner, it suffices to maximize only over the Werner states, rather than maximizing over \emph{all} states, with negativity at most~$c$. Analogously, when $\tau=\eta_\beta^d$ is isotropic, it suffices to maximize only over the isotropic states. Explicit calculations to evaluate the support function $h_{\calN_c}$ on Werner and isotropic states are given Appendix~\ref{app:wernerisotropiccalculations}.

Using these calculations, computable witnesses can be defined by optimizing over the isotropic and Werner states. These are
\[
 W_{\text{wer}}(\rho,\sigma): = d^2\, \min_{\tau\text{ Werner}} W_{\tau}(\rho,\sigma)
\]
and
\[
 W_{\text{iso}}(\rho,\sigma): = d^2\, \min_{\tau\text{ isotropic}} W_{\tau}(\rho,\sigma).
\]
Explicit calculations can again be found in Appendix \ref{app:Werner}, but closed-form results for this witnesses can be given as
\begin{equation}\label{eq:wernerwitness}
 W_{\text{wer}}(\rho,\sigma)=  \min\left\{W_N(\rho,\sigma),\,  \tfrac{2}{d(d-1)}W_{\text{wer}}'(\rho,\sigma) \right\},
\end{equation}
and
\begin{equation}\label{eq:isotropicwitness}
 W_{\text{iso}}(\rho,\sigma)=  \min\left\{W_N(\rho,\sigma),\, W_{\text{iso}}'(\rho,\sigma) \right\},
\end{equation}
where $W'_{\text{iso}}$ and $W'_{\text{wer}}$ are sub-witness given by
\begin{equation}\label{eq:wernersubwitness}
 W_{\text{wer}}'(\rho,\sigma) = \frac{dN(\tilde\rho)+1}{2}N(\rho)-\Tr[F_-\sigma^\Gammaminus],
\end{equation}
and
\begin{equation}\label{eq:isotropicsubwitness}
 W_{\text{iso}}'(\rho,\sigma) = \frac{2N(\tilde\rho)+1}{d}N(\rho)-\bra{\Phi}\sigma^\Gammaminus\ket{\Phi}.
\end{equation}
Here $F_-$ is the negative part of the flip operator.

From the expressions for the witnesses in~\eqref{eq:wernerwitness} and~\eqref{eq:isotropicwitness}, it is clear that $W_{\text{iso}}(\rho,\sigma)\leq W_N(\rho,\sigma)$ for all $\rho$ and $\sigma$, and thus $W_\text{iso}\succeq W_N$. Similarly, we have \mbox{$W_\text{wer}\succeq W_N$}. So both new witnesses do in fact supply at least as much information about convertibility of states as the negativity does. 

However, in the next subsection, we show that all of the witnesses $W_N$, $W_\text{wer}'$ and $W'_\text{iso}$ are incomparable with respect to the partial order on witnesses.  Minimizing over all three of these witnesses 
\begin{equation}\label{eq:Wgamma}
 W_\Gamma=\min\left\{W_N,\,W'_\text{wer},\,W'_\text{iso}\right\}
\end{equation}
yields a computable witness that is certainly an improvement over the negativity. 

Interestingly, the witnesses in \eqref{eq:wernersubwitness} and \eqref{eq:isotropicsubwitness} do not depend on the explicit form of $\rho$, but only on the negativities $N(\rho)$ and $N(\tilde\rho)$. Furthermore, note that $\bra{\Phi}\sigma^\Gammaminus\ket{\Phi}\leq \Tr[\sigma^\Gammaminus]=N(\sigma)$, and thus
\begin{equation*}
 \frac{2N(\tilde\rho)+1}{d}N(\rho)-N(\sigma)\leq W'_{\text{iso}}(\rho,\sigma).
\end{equation*}
Therefore, this new witness $W_{\text{iso}}$ can only supply new information about whether or not $\rho\mapsto\sigma$ if $N(\tilde\rho)<\frac{d-1}{2}$.  
That is, if $N(\rho)\geq N(\sigma)$ but $W_\text{iso}(\rho,\sigma)<0$ for some states $\rho$ and $\sigma$, then we must have that $N(\tilde\rho)<\frac{d-1}{2}$.

Similarly, note that $\Tr[F_-\sigma^\Gammaminus]\leq N(\sigma)$ and thus 
\[
 \frac{dN(\tilde\rho)+1}{2}N(\rho)-N(\sigma)\leq W'_{\text{wer}}(\rho,\sigma),
\]
so the new witness $W_{\text{wer}}$ can only supply new information about whether or not $\rho\mapsto\sigma$ if $N(\tilde\rho)<\frac{1}{d}$.

\subsection{Effectiveness in two-qubit systems}

We first analyze the effectiveness of our new conversion witnesses when both $\rho$ and $\sigma$ are states of a two-qubit system, in which case the two witnesses are actually the same. The witness is compared against the negativity and the entanglement of formation, two known entanglement monotones that can be computed for all states of two qubits. The negativity and entanglement of formation are inequivalent as witnesses, since they detect different orderings of two-qubit states \cite{Miranowicz2004}. Here we show that our witness is better than both of these monotones in some cases.

The entanglement of formation for two-qubit states is related to the \emph{concurrence} by
\[
 E(\rho)=H\left(\frac{1+\sqrt{1-C(\rho)^2}}{2}\right),
\]
where $H(x)$ is the binary entropy function. Here $C(\rho)$ is the concurrence, which is defined only for two-qubit states and is an entanglement monotone in its own right \cite{Wootters1998}. For pure states of the form $\ket{\psi}=\sum_{i,j=0}^1\psi_{ij}\ket{ij}$, the concurrence is defined by $C(\ketbra{\psi}{\psi})=\abs{\psi_{00}\psi_{11}-\psi_{01}\psi_{10}}$ and can be extended to mixed states by
\[
 C(\rho)=\max\{0,\, \sqrt{\mu_1}-\sqrt{\mu_1}-\sqrt{\mu_1}-\sqrt{\mu_1}\}
\]
where the $\mu_i$ are the eigenvalues (in non-increasing order) of the matrix $\sigma_y\otimes\sigma_y\rho\sigma_y\otimes\sigma_y\overline{\rho}$, where $\sigma_y=\left(\begin{smallmatrix}0&-i\\i&0\end{smallmatrix}\right)$ is the Pauli matrix and $\overline{\rho}$ is the matrix whose entries are the complex conjugates of $\rho$.

We now compare our new witness to the concurrence and the negativity. In particular, there exist families of two-qubit states for which our witness performs better than both the concurrence and the negativity at detecting non-convertibility. Our witness is not always better than the concurrence, since we also present pairs of states for which the concurrence detects non-convertibility but our witness does not.

With $d=2$, note that $F_-=\ketbra{\psi^-_{01}}{\psi^-_{01}}$ where 
\[\ket{\psi^-_{01}}=\tfrac{1}{\sqrt{2}}\left(\ket{01}-\ket{10}\right).\]
This state is related to the standard maximally tangled state $\ket{\Phi}=\frac{1}{\sqrt{2}}\left(\ket{00}+\ket{11}\right)$ by local unitary matrices, so the witnesses $W_\text{wer}$ and $W_\text{iso}$ can be reduced to a single equivalent conversion witness
\begin{equation}\label{eq:2qwit}
 W(\rho,\sigma)=\Tr[\rho^{\Gammaminus\Gammaminus}]+\frac{1}{2}N(\rho)-\Tr[F_-\sigma^{\Gammaminus}]
\end{equation}
in the case when both $\rho$ and $\sigma$ are two-qubit states. 

Furthermore, if $\sigma=\ketbra{x}{x}$ is a pure state of the form $\ket{x}=\sqrt{\lambda_0}\ket{00}+\sqrt{\lambda_1}\ket{11}$ then $\sigma^{\Gammaminus}=\sqrt{\lambda_0\lambda_1}\ketbra{\psi^-_{01}}{\psi^-_{01}}$. In this case the conversion witness in \eqref{eq:2qwit} further reduces to
\[
 W(\rho,\sigma) = \Tr[\rho^{\Gammaminus\Gammaminus}]+\frac{1}{2}N(\rho)-\sqrt{\lambda_0\lambda_1}
\]
where $N(\ketbra{x}{x})=\sqrt{\lambda_0\lambda_1}$.

Let $\rho=\rho_q$ be in the family of states given by
\[
 \rho_q = \frac{1}{2}\left(\ketbra{\Phi}{\Phi}+\ketbra{\psi_q}{\psi_q}\right)
\]
parametrized by $0\leq q\leq 1$, where
\[
 \ket{\psi_q}=\sqrt{q}\ket{00}+\sqrt{1-q}\ket{10}.
\]
All of the states in this family have the same concurrence $C(\rho_q)=\frac{1}{2}$ \cite{Miranowicz2004}, so they all have the same entanglement of formation as well. The negativities of these states are
\[
 N(\rho_q)=\frac{\sqrt{2(1+q)}-1}{4}.
\]
Let $\sigma=\ketbra{x}{x}$ be a pure state parametrized by the largest Schmidt coefficient $\lambda_0$.

\begin{figure}
 \centering
 \includegraphics[width=\columnwidth]{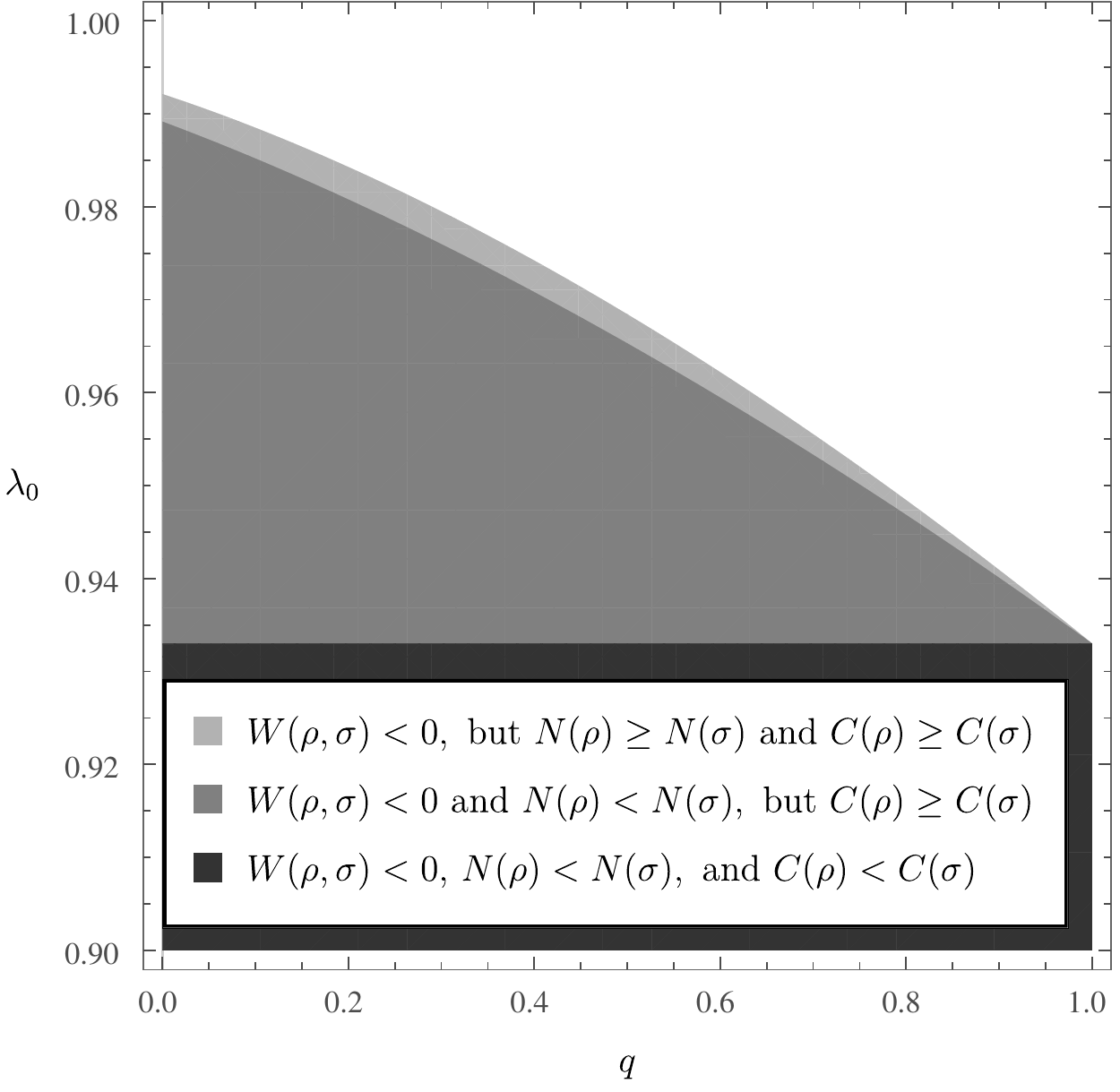}
 \caption{Comparison of the new conversion witnesses $W(\rho,\sigma)$ against both the negativity and concurrence for states $\rho$ and $\sigma$ of two qubits, where $\rho=\rho_q$ and $\sigma=\ketbra{x}{x}$ as defined in the main body of the text. The lightly shaded strip near the top of the figure denotes pairs of states for which our witness is better than \emph{both} the negativity and the concurrence at detecting non-convertibility.}
 \label{fig:twoqubitcompare}
\end{figure}

For the families of states $\rho$ and $\sigma$ given above, a comparison of our new conversion witness $W$ against both the negativity and concurrence can be seen in FIG.\ \ref{fig:twoqubitcompare}. The lightly shaded strip near the top of the figure denotes pairs of states for which our witness detects that $\rho\not\mapsto\sigma$, but the negativity and the concurrence do not. 

Since the concurrence as an entanglement monotone is equivalent to the entanglement of formation, the examples above show that our new conversion witness can detect the non-convertibility of some two-qubit state transformations that cannot be determined from either the negativity or the entanglement of formation. Our witness therefore allows us to see more of the structure of the ordering of entangled two-qubit states that was previously possible. But our witness is not always better than these monotones. If $\rho$ and $\sigma$ are both pure states of two qubits, for example, then $\Tr[\rho^{\Gammaminus\Gammaminus}]=\frac{1}{2}N(\rho)$ and the conversion witness reduces to
\[
 W(\rho,\sigma)=N(\rho)-N(\sigma),
\]
which is just the difference of negativities. 

We now present states $\rho$ and $\sigma$ for which the concurrence detects that $\rho\not\mapsto\sigma$ while our witness does not. Let $\rho=\omega_\alpha$ be the Werner state of two qubits with $\alpha=\frac{1-\sqrt{2}}{2}$, and let
\[
 \sigma=\tfrac{1}{2}\ketbra{\Phi}{\Phi} + \tfrac{1}{2}\ketbra{01}{01}.
\]
These states have the same negativity, which is $N(\rho)=N(\sigma)=\frac{\sqrt{2}-1}{4}$, but different values of the concurrence,
\[
 C(\rho)=\frac{\sqrt{2}-1}{2}\qquad \text{and}\qquad C(\sigma)=\frac{1}{2}.
\]
Since $C(\rho)<C(\sigma)$, the concurrence detects the inconvertibility of $\rho\not\mapsto\sigma$ while the negativity does not. Furthermore, since $\Tr[\rho^{\Gammaminus\Gammaminus}]=\frac{\sqrt{2}-1}{8}$ and $\Tr[\sigma^{\Gammaminus}F_-]=\frac{1}{8\sqrt{2}}$, our conversion witness evaluates to 
\[
 W(\rho,\sigma)=\frac{3\sqrt{2}-4}{16}>0,
\]
so our witness does not detect this inconvertibility. On the other hand, since $C(\sigma)>C(\rho)$ we cannot determine from the concurrence alone if the reverse conversion $\sigma\mapsto\rho$ is possible. Since $\Tr[\sigma^{\Gammaminus\Gammaminus}]=\frac{2-\sqrt{2}}{16}$ and $\Tr[\rho^{\Gammaminus}F_-]=\frac{\sqrt{2}-1}{4}$, our conversion witness evaluates to
\[
 W(\sigma,\rho) = \frac{4-3\sqrt{2}}{16} <0.
\]
Our witness does detect the non-convertibility of $\sigma\not\mapsto\rho$, while the concurrence and negativity do not.

Neither the concurrence nor our new conversion witness is strictly better than the other at determining which state conversions possible. In the framework of the partial ordering of no-go entanglement conversion witnesses, this means that $W\not\succeq W_C$ and $W\not\preceq W_C$, where $W_C$ is the no-go conversion witness defined by the concurrence
\[
 W_C(\rho,\sigma) = C(\rho)-C(\sigma)
\]
for two-qubit states. Although it does not completely supersede the importance of the entanglement of formation, the new conversion witness does improve our knowledge about which state transformations are possible in systems of two qubits.

\subsection{Incomparability of Conversion Witnesses}
\label{sec:improvement}

For states of systems that are larger than two qubits the entanglement of formation can no longer be computed. The negativity is the only known computable entanglement monotone against which we can compare the effectiveness of our new no-go conversion witness, and we show here that our witness is actually better at detecting non-convertibility. It was shown in section \ref{sec:computablenogowitness} that $W_\text{wer}$ and $W_{\text{iso}}$ in \eqref{eq:wernerwitness} and \eqref{eq:isotropicwitness} give us at least as much information as the negativity, but it is not obvious that they are actually an improvement. In this section, we present families of pairs of states ($\rho$ and $\sigma$) to illustrate that $W_N\not\succeq W_\text{iso}$ and $W_N\not\succeq W_\text{wer}$.  In addition, we also show that neither of the new witnesses $W_\text{wer}$ and $W_\text{iso}$ are strictly better than the other.

\begin{proposition}
 The witnesses $W_N$, $W_\textup{wer}'$ and $W'_\textup{iso}$ are all incomparable with respect to the partial order. That is,
 \begin{gather*}
  W_N\not\succeq W'_\textup{wer},W'_\textup{iso}, \hspace{5mm} W'_\textup{wer}\not\succeq W_N,W'_\textup{iso}, \\\text{and}\hspace{3mm} W'_\textup{iso}\not\succeq W_N,W'_\textup{wer}.
 \end{gather*}

\end{proposition}
Hence the no-go witness $W_\Gamma$ obtained in \eqref{eq:Wgamma} from minimizing over all three of the above witnesses truly does give more information about the conversion of states under PPT than the negativity.

This proposition is proved by finding pairs of states $\rho$ and $\sigma$ where one witness detects the inconvertibility $\rho\not\mapsto\sigma$ while the other two do not, Hence none of these three witnesses is greater than any other with respect to the partial order.

\subsubsection{Proof that \texorpdfstring{$W'_\textup{iso}$}{Wiso} is not less than \texorpdfstring{$W_N$}{WN} or \texorpdfstring{$W'_\textup{wer}$}{Wwer}}

We first show that $W_N\not\succeq W'_\text{iso}$ and $W'_\text{wer}\not\succeq W'_\text{iso}$. That is, there are states $\rho$ and $\sigma$ such that $W'_\text{iso}(\rho,\sigma)<0$, but $W_N(\rho,\sigma)\geq0$ and $W'_\text{wer}(\rho,\sigma)\geq0$.   In a bipartite $d\times d$-system, such an example can only exist if $d\geq3$.

Let $d\geq3$ and consider $\sigma$ to be a Werner state of a $d\times d$-system as given in~\eqref{eq:wernerparameter}.  The negativity of the Werner states is given by
\[
 N(\omega_\alpha^d)=\left\{\begin{array}{ll}
                       \frac{-\alpha}{d}, & -1\leq \alpha < 0\\
                       0, & 0\leq \alpha\leq 1.
                      \end{array}
\right.
\]
Note that the Werner state $\omega_\alpha^d$ is entangled if and only if $N(\omega_\alpha^d)>0$. In this case, we have
\[
 (\omega_\alpha^d)^\Gammaminus = N(\omega_\alpha^d)\ketbra{\Phi}{\Phi}
\]
and note that the most entangled Werner state has negativity $\frac{1}{d}$ with $\alpha=-1$.

Let $\rho=\ketbra{x}{x}$ be a pure state of two qubits, which we may consider to be in Schmidt form
\[
 \ket{x}=\sqrt{\lambda_0}\ket{00}+\sqrt{\lambda_1}\ket{11}
\]
where $\lambda_0\geq\lambda_1\geq0$ are the Schmidt coefficients such that $\lambda_0+\lambda_1=1$. This state has negativity $N(\ketbra{x}{x}) = \sqrt{\lambda_0\lambda_1}$. Note that $\ketbra{x}{x}^\Gammaminus=\sqrt{\lambda_0\lambda_1}\ketbra{\psi_{01}^-}{\psi_{01}^-}$. For $\rho=\ketbra{x}{x}$, this yields $\tilde\rho=\frac{1}{N(\rho)}\rho^\Gammaminus= \ketbra{\psi_{01}^-}{\psi_{01}^-}$ and thus $N(\tilde\rho)=\frac{1}{2}$. 

With states $\rho=\ketbra{x}{x}$ and $\sigma=\omega_\alpha^d$, the witness in \eqref{eq:isotropicsubwitness} evaluates to
\begin{align*}
 W'_{\text{iso}}(\ketbra{x}{x},\omega_\alpha^d) &= \frac{2N(\ketbra{x}{x})}{d} - N(\omega_\alpha^d)\\ 
 &= \frac{1}{d}\left(2\sqrt{\lambda_0\lambda_1}+\alpha\right).
\end{align*}
Our new witness therefore shows that the conversion $\ketbra{x}{x}\mapsto \omega_\alpha^d$ is not possible if $2\sqrt{\lambda_0\lambda_1}+\alpha<0$. From the witness given by the difference in negativities, we can determine that this conversion is not possible when $N(\ketbra{x}{x})<N(\omega_\alpha)$, which occurs when $d\sqrt{\lambda_0\lambda_1} +\alpha<0$. As long as $d>2$, these witnesses are clearly distinct for these families of states. 

\begin{figure}
 \centering
 \includegraphics[width=\columnwidth]{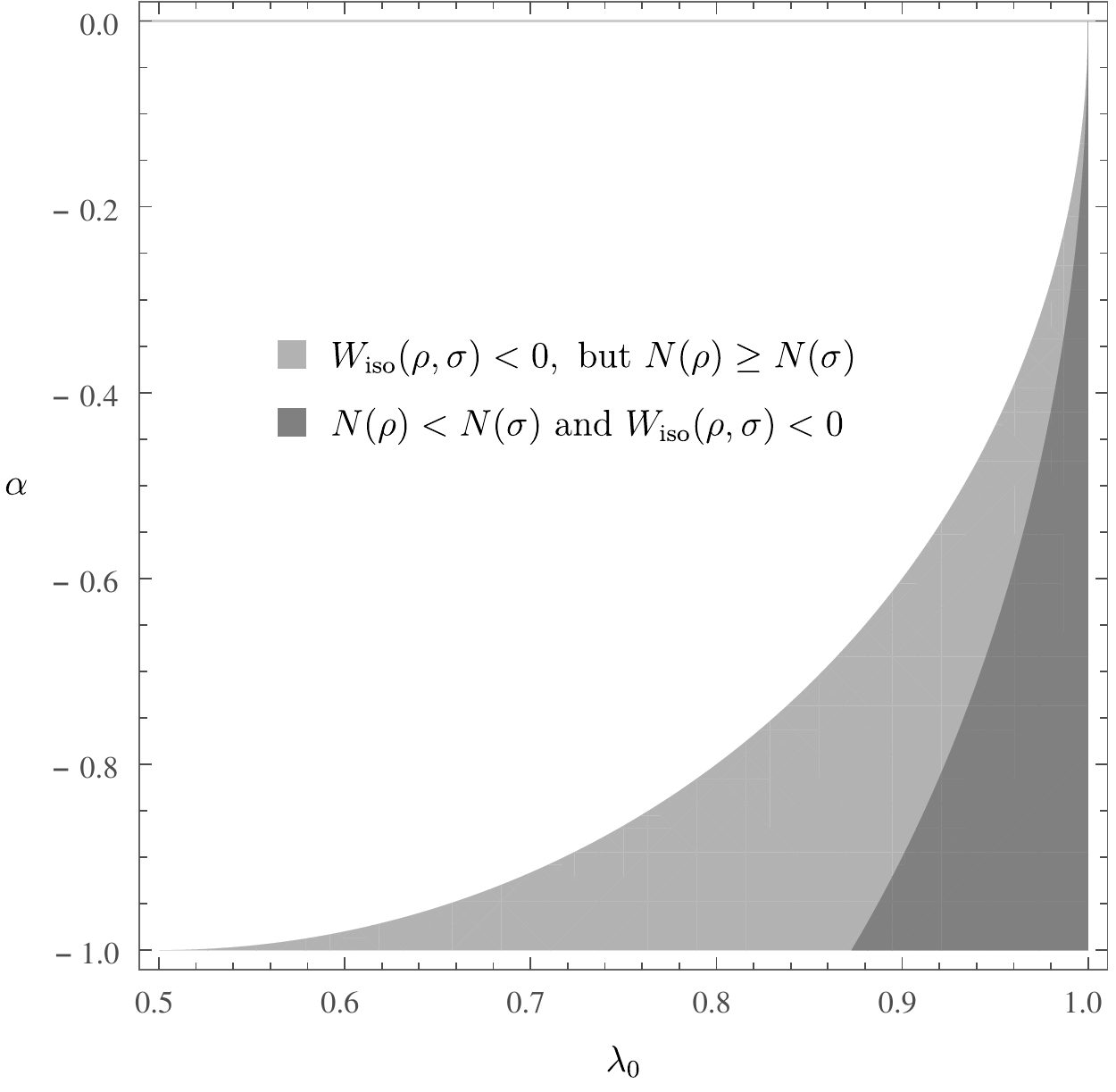}
 \caption{Comparison of the witnesses $W_N(\rho,\sigma)$ and $W_\text{iso}(\rho,\sigma)$  for states $\rho=\ketbra{x}{x}$, with $\ket{x}=\sqrt{\lambda_0}\ket{00}+\sqrt{\lambda_1}\ket{11}$, and Werner states $\sigma=\omega_\alpha$ with $d=3$. The largest Schmidt coefficient $\lambda_0$ ranges from~$\frac{1}{2}$ (maximally entangled state) to $1$ (separable state) on the horizontal axis while the Werner parameter $\alpha$ ranges from $-1$ to $0$ on the vertical axis. The lightly shaded region denotes pairs of states for which $W_\text{iso}$ detects non-convertibility but $W_N$ does not.}
 \label{fig:wisovneg3}
\end{figure}

For $d=3$ the distinction between $W_\text{iso}(\rho,\sigma)$ and $W_N(\rho,\sigma)$ is depicted in FIG. ~\ref{fig:wisovneg3}. Points inside the shaded regions denote pairs of states for which the conversion $\rho\mapsto\sigma$ is not possible. In the darker region, both $W_\text{iso}$ and $W_N$ detect the non-convertibility while only $W_\text{iso}$ does so in the lightly region.  

The diagram in FIG.\ \ref{fig:wisovneg3} shows that this witness really does provide more information than the negativity. In particular, we can choose 
\[
   \lambda_0=\frac{d+\sqrt{d^2-4}}{2d}\hspace{5mm}\text{and}\hspace{5mm} \lambda_1=\frac{d-\sqrt{d^2-4}}{2d}
\] 
such that $N(\ketbra{x}{x})=\frac{1}{d}$ and $\sigma=\omega_\alpha$ with $\alpha=-1$ such that $N(\sigma)=\frac{1}{d}$. Thus $\rho$ and $\sigma$ have the same negativity, i.e.\ $W_N(\rho,\sigma)\geq0$, so we cannot determine from the negativity alone if $\rho\mapsto\sigma$. However, the no-go witness $W'_{\text{iso}}$ for these states evaluates to 
\[
 W'_{\text{iso}}(\rho,\sigma)=\frac{2}{d^2}-\frac{1}{d} <0,
\]
and we conclude that $\rho\not\mapsto\sigma$. Hence $W_N\not\succeq W'_\text{iso}$.  

Furthermore, note that $\Tr[\ketbra{\Phi}{\Phi}F_-]=0$ and thus $W'_\text{wer}(\rho,\sigma)\geq0$ for the above choice of $\rho$ and $\sigma$. This shows that $W'_\text{wer}\not\succeq W'_\text{iso}$.

\subsubsection{Proof that \texorpdfstring{$W_\textup{wer}$}{Wwer} is not less than \texorpdfstring{$W_N$}{WN} or \texorpdfstring{$W_\textup{iso}$}{Wiso}}

Consider a family of pure states $\rho=\ketbra{y}{y}$ of two qutrits, where $\ket{y}$ may be given in Schmidt form as
\[
 \ket{y}=\sqrt{\lambda_0}\ket{00}+\sqrt{\lambda_1}\ket{11}+\sqrt{\lambda_2}\ket{22},
\]
with $\lambda_0=\tfrac{9}{10}$, $\lambda_2=\frac{1}{10}-\lambda_1$, and $\frac{1}{20}\leq\lambda_1\leq\frac{1}{10}$. The negativity of the state $\rho=\ketbra{y}{y}$ is 
\begin{align*}
 N(\rho) &= \sqrt{\lambda_0\lambda_1}+\sqrt{\lambda_0\lambda_2}+\sqrt{\lambda_1\lambda_2}.
\end{align*}
Let $\sigma=\eta_{\beta}$ be an isotropic state of a $2\times 2$ system. The negative component of the flip operator with $d=2$ is simply $F_-=\ketbra{\psi^-_{01}}{\psi^-_{01}}$, so $\Tr[\sigma^\Gammaminus F_-] = N(\eta_\beta)=\frac{2\beta-1}{2}$. Evaluating the witness $W_\text{wer}(\rho,\sigma)$ on these states yields
\begin{equation}\label{eq:wwerex}
 W_\text{wer}(\rho,\sigma)=\frac{2\Tr[\rho^{\Gammaminus\Gammaminus}]+N(\rho)}{2} - N(\sigma),
\end{equation}
which is clearly different from the difference of negativities $W_N(\rho,\sigma)=N(\rho)-N(\sigma)$.

\begin{figure}
 \centering
 \includegraphics[width=\columnwidth]{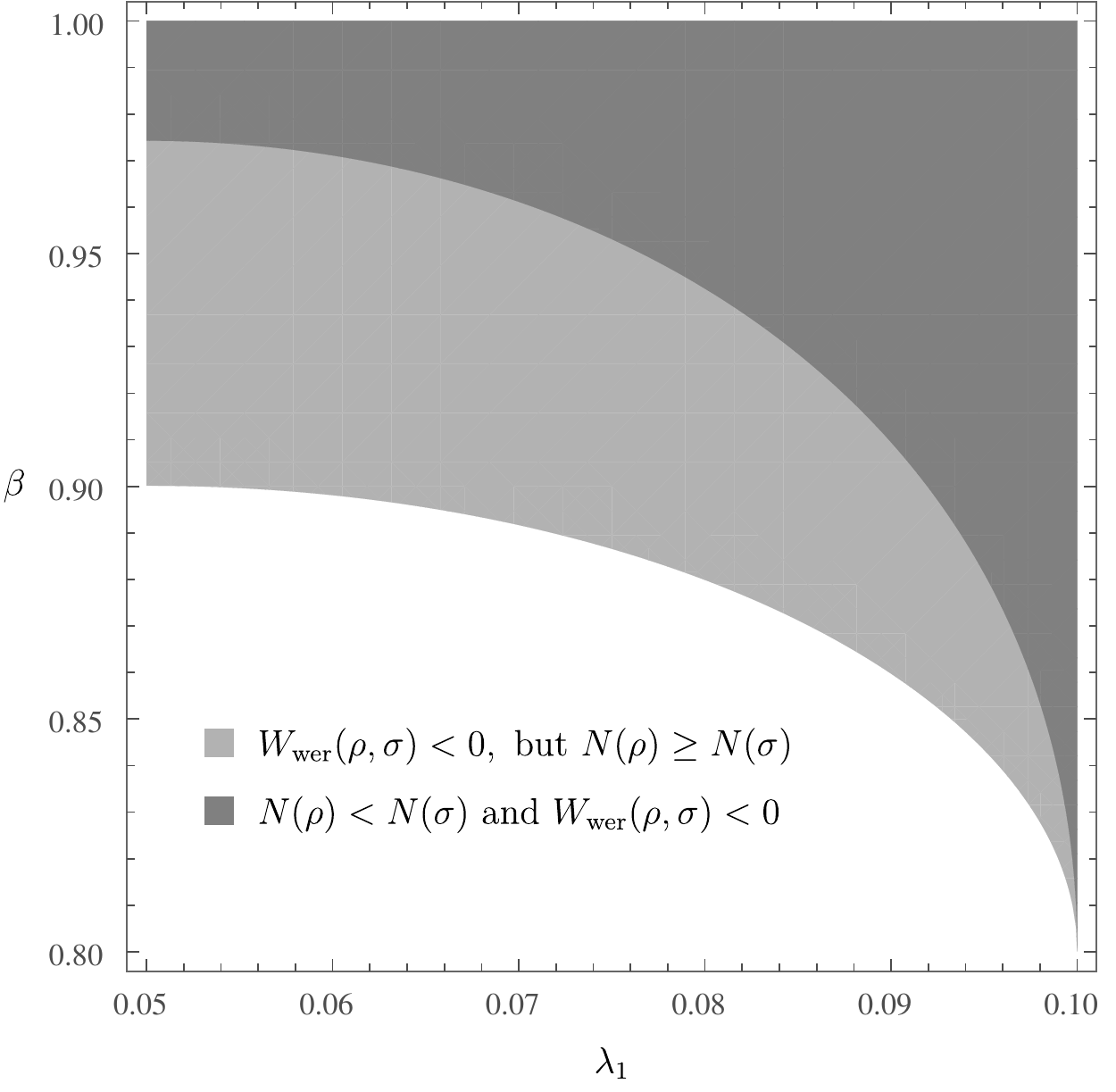}
 \caption{Comparison of the witnesses $W_N(\rho,\sigma)$ and $W_\text{wer}(\rho,\sigma)$  for states $\rho=\ketbra{y}{y}$, with $\ket{y}=\sqrt{\frac{9}{10}}\ket{00}+\sqrt{\lambda_1}\ket{11}+\sqrt{\frac{1}{10}-\lambda_1}\ket{22}$, and isotropic states $\sigma=\eta_\beta$ with $d=2$. The Schmidt coefficient $\lambda_1$ ranges from~$\frac{1}{20}$ to $\frac{1}{10}$ on the horizontal axis while the isotropic parameter $\beta$ ranges from $0.8$ to $1$ on the vertical axis. The lightly shaded region denotes pairs of states for which $W_\text{iso}$ detects non-convertibility but $W_N$ does not.}
 \label{fig:wwervneg}
\end{figure}

The distinction between $W_\text{wer}(\rho,\sigma)$ and $W_N(\rho,\sigma)$ for the families of states $\rho$ and $\sigma$ given above is depicted in FIG.\ \ref{fig:wwervneg}. Points inside the shaded regions denote pairs of states for which the witnesses $W_\text{wer}$ and $W_N$ detect that the conversion $\rho\mapsto\sigma$ is not possible. In the darker region, both $W_\text{wer}$ and $W_N$ detect the non-convertibility while only $W_\text{wer}$ does so in the lightly region.

The diagram in FIG.\ \ref{fig:wwervneg} shows that this witness really does provide more information than the negativity. In particular, we can choose $\lambda_1=\frac{1}{20}$, and thus $\lambda_2=\lambda_1$, such that the negativity of the state $\rho=\ketbra{y}{y}$ is 
\begin{align*}
 N(\rho) &= \frac{6\sqrt{2}+1}{20} \approx 0.474.
\end{align*}
Hence the negativity detects that $\rho$ cannot be converted into an isotropic state $\sigma=\eta_\beta$ with $d=2$ when $N(\rho)<N(\sigma)$ or
\[
 \beta>N(\rho)+\frac{1}{2}\approx0.974.
\]
Now we determine the values of $\beta$ for which $W_\text{wer}$ detects this non-convertibility. It is straightforward to compute that $\Tr[\rho^{\Gammaminus\Gammaminus}]=\frac{1+\sqrt{145}}{80}$. From \eqref{eq:wwerex}, the witness $W_\text{wer}$ evaluates to
\begin{equation*}\label{eq:b}
 W_\text{wer}(\rho,\sigma) 
 = \frac{3+12\sqrt{2}+\sqrt{145}}{80} -N(\sigma).
\end{equation*}
The witness $W_\text{wer}$ detects non-convertibility exactly when $W_\text{wer}(\rho,\sigma)<0$, or
\begin{equation*}
 \beta> \frac{43+12\sqrt{2}+\sqrt{145}}{80} \approx 0.900.
\end{equation*}
Hence our new witness $W_\text{wer}$ is better than the negativity at detecting that $\rho\not\mapsto\sigma$ for these states when $\beta$ is in the range $0.900<\beta<0.974$. This range is seen on the left-hand side of the lightly shaded are of FIG.\ \ref{fig:wwervneg}. We conclude that  $W_N\not\succeq W'_\text{wer}$.

Furthermore, note that $\Tr[\ketbra{\Phi}{\Phi}F_-]=0$ and thus $W'_\text{iso}(\rho,\sigma)\geq 0$ for the above choice of $\rho$ and $\sigma$. This shows that $W'_\text{iso}\not\succeq W'_\text{wer}$.

\subsubsection{Proof that \texorpdfstring{$W_N$}{WN} is not less than \texorpdfstring{$W_\textup{wer}$}{Wwer} or \texorpdfstring{$W_\textup{iso}$}{Wiso}}

Let $\rho$ be any unentangled state. Let $d\geq2$ and let $\sigma$ be any entangled Werner state of a $d\times d$-dimensional system. Note that $\Tr[F_-\sigma^\Gammaminus]=0$ since $\sigma$ is a Werner state, and thus $W'_\text{wer}(\rho,\sigma)\geq 0$. However, from the negativities, we see that $N(\sigma)>0$ but $N(\rho)=0$. Hence $W_N(\rho,\sigma)<0$ and thus $\rho\not\mapsto\sigma$. Therefore $W'_\text{wer}\not\succeq W_N$. 

As above, let $\rho$ be any unentangled state and $d\geq 2$. Now, however, let $\sigma$ be any entangled isotropic state of a $d\times d$-dimensional system. Note that $\Tr[\ketbra{\Phi}{\Phi}\sigma^\Gammaminus]=0$ since $\sigma$ is isotropic, and thus $W'_\text{iso}(\rho,\sigma)\geq 0$. However, from the negativities we have that $W_N(\rho,\sigma)<0$ and $\rho\not\mapsto\sigma$. Hence $W'_\text{iso}\not\succeq W_N$. 

\subsection{Further PPT conversion witnesses}

In this section, we show how further conversion witnesses might be constructed that are better than those presented in the preceding section.

While the witnesses constructed above are in fact computable and are indeed an improvement over the negativity, these quantities have the unfortunate property that they are not invariant under local transformations. That is, for local unitaries $U$ and $V$, we don't necessarily have that 
\[
 W(\rho, U\otimes V \sigma U^\dagger \otimes V^\dagger) = W(\rho,\sigma),
\]
In principle, since the application of local unitaries is an invertible operation in the class of LOCC (and PPT) operations, we should have
\[
\rho\mapsto \sigma  \hspace{ 5mm}\Longleftrightarrow \hspace{4mm}  \rho\mapsto U\otimes V \sigma U^\dagger \otimes V^\dagger
\]
for any unitaries $U$ and $V$. That is, if $\rho$ can be converted to $\sigma$, then it can also be converted into any state that is equivalent to $\sigma$ up to local unitaries. Hence, if $W$ is a no-go witness and there exist some $U$ and $V$ such that $W(\rho, U\otimes V \sigma U^\dagger \otimes V^\dagger)<0$, then $\rho\not\mapsto\sigma$. 

Since our witnesses constructed in the previous sections depend on a particular choice of bases relative to the two systems, this suggests that we might be able to build an improved witness by minimizing over all local unitaries (LU):
\begin{equation}
\label{eq:LUwitness}
 W^\text{LU}(\rho,\sigma) = \min_{U,V} W(\rho, U\otimes V \sigma U^\dagger \otimes V^\dagger).
\end{equation}
where $W$ is either $W_\text{iso}$ or $W_\text{wer}$. This extension of these witnesses makes them no longer directly computable. Yet, in contrast to the computation of the support function $h_{\calN(\tilde\rho)}$ for arbitrary $\rho$, the minimization in~\eqref{eq:LUwitness} can be performed numerically, so perhaps not all is lost. In particular, the witness defined by
\[
 W^\text{LU}_\Gamma (\rho,\sigma) = \min_{U,V}  W_\Gamma(\rho,U\otimes V \sigma U^\dagger \otimes V^\dagger)
\]
might be useful for determining convertibility of quantum states within the context of entanglement theory.

The witnesses $W_\text{wer}$ and $W_\text{iso}$ that we have constructed here are computable because the support function $h_{\calN_c}(\tau)$ is computable when $\tau$ is a Werner or isotropic state. This computability is due to the high degree of symmetry present in these families of states which comes from the fact that they are invariant under a particular group action. However, there are other group actions that one might be able to consider, through which other families of highly symmetric states could be constructed where the support function $h_{\calN_c}$ is also computable. Indeed, if $G$ is a group with an action on states given by $U(g)\rho U^\dagger(g)$ for $g\in G$, we can make use of a twirling operation $\calT_G(\rho)=\int_G U(g)\rho U^\dagger(g)\, dg$, where $dg$ is the Haar measure of $G$. If this twirling operation is a PPT operation and $\tau$ is invariant under the action of $G$, then $\sup_{\gamma\in\calN_c}\Tr[\gamma\tau]=\Tr[\gamma^\star\tau]$ will be maximized by state $\gamma^\star$ that is invariant under $G$. Hence, finding a group $G$ with a suitable group action could lead to the construction of other computable conversion witnesses.

For example, the Werner states are invariant under all bipartite unitaries of the form $U\otimes U$, where $U$ is any $d\times d$ unitary matrix. But the Werner states can be considered as a special class of states of the form
\begin{equation}\label{eq:genwernerstates}
 \rho = a \sum_{i}\ketbra{ii}{ii} + b\sum_{i<j}\ketbra{\psi_{ij}^+}{\psi_{ij}^+} + c\sum_{i<j}\ketbra{\psi_{ij}^-}{\psi_{ij}^-},
\end{equation}
which is the family of states that are invariant under application of bipartite unitaries of the form $U\otimes U$ where~$U$ is diagonal in the $\{\ket{i}\}$ basis with
\[                                                                                                                                        
  \bra{i}U\ket{j}=  \delta_{ij}e^{i\theta_j}.
\]
We shall call the states in \eqref{eq:genwernerstates} the generalized Werner states. This family of states has been studied before in relation to the entanglement distillation problem~\cite{DiVincenzo2000}. Making use of the symmetry in these states, it is also possible to compute $h_{\calN_c}(\tau)$ whenever $\tau$ is a generalized Werner state. However, minimizing $W_\tau(\rho,\sigma)$ over all of the generalized Werner states turns out to be no less then $W_\text{wer}$, which is obtained from minimizing over just the Werner states, so this particular family of symmetric states does not yield any improved witness (see Appendix \ref{app:rhoabstates} for calculations). This indicates that the witness $W_\text{wer}$ that we have constructed above is in fact quite good, since its generalization is not any better.

Finally, we show how one can construct refinements of our no-go witnesses that are not necessarily computable. Although they are not computable, they may yield insight into the problem of constructing other useful witnesses for PPT conversion. Note that key idea in constructing the PPT conversion witness in the preceding sections was that a state $\sigma$ cannot be obtained from $\rho$ if  $\sigma\not\in\calC_\Gamma(\rho)$, i.e.\ if $\sigma$ is outside of the PPT-orbit of $\rho$. But each state that can be obtained from $\rho$ must have negativity at most $N(\rho)$. The essential ingredient in the construction of our computable no-go conversion witness was thus the approximation of the PPT-orbit by the set of states $\calN_{c}$ with negativity at most $c=N(\rho)$, where we note that $\calC_\Gamma(\rho)\subset \calN_{N(\rho)}$. This observation, however, only yields the information that $N(\sigma)>N(\rho)$ (and thus $\sigma\not\in\calC_\Gamma(\rho)$) implies that $\sigma$ can't be obtained from $\rho$. 

Our witness makes a stronger statement than this. In particular, we have that 
\[
 \sigma^{\Gammaminus}\not\leq \gamma \hspace{5mm}\text{for all }\gamma\in \calC_\Gamma(\rho^\Gammaminus)
\]
implies that $\rho\not\mapsto\sigma$. Thus,
\[W_\tau(\rho,\sigma) = h_{\calC_\Gamma(\rho^\Gammaminus)}(\tau)-\Tr[\tau\sigma^\Gammaminus]\]
is a witness for each $\tau$. But the value of the support function $h_{\calC_\Gamma(\rho^\Gammaminus)}$ cannot be found in general, since  minimizing over the PPT-orbit of $\rho^\Gammaminus$ is difficult. Instead, we approximated $\calC_\Gamma(\rho^\Gammaminus)$ with the larger set
\[
 \calN_1(\rho)=\big\{\gamma\in H_{n,+}  \big| \Tr[\gamma]= N(\rho), \, \Tr[\gamma^\Gammaminus]<\Tr[\rho^{\Gammaminus\Gammaminus}] \big\}
\]
whose support function $h_{\calN_1(\rho)}(\tau)$ is in fact computable for certain $\tau$, in particular for Werner and isotropic states. Note that the set $\calN_1(\rho)$ is precisely the unnormalized version of the states in the set $\calN_{N(\tilde\rho)}$ in \eqref{eq:negsets}. Yet even this approximation can be further refined. For sets $\calS$ that are refinements of $\calN_1(\rho)$, i.e. such that $\calC_\Gamma(\rho^\Gammaminus)\subset \calS\subset \calN_1(\rho)$, we can construct a witness 
\[
 W_{\calS,\tau}(\rho,\sigma) = h_{\calS}(\tau) -\Tr[\tau\sigma^\Gammaminus]
\]
for each state $\tau$. The derivation of the refinements $\calS$ are left to Appendix \ref{app:further}, since the details are rather cumbersome and may not lead to an improved witness which is computable.


\section{Conclusion}
\label{sec:conclusion}

Although the concept of conversion witnesses in quantum resource theories was first introduced elsewhere, we have demonstrated the importance of studying conversion witnesses within the framework of entanglement theory. The central problem in entanglement theory is to determine conditions for when one state may be converted into another. Such conditions have, up until now, only been studied in terms of monotones. We have shown that conversion witnesses, which may be considered as a generalization of monotones, can be used to determine convertibility of quantum states when the best-known monotones fail to do so. Moreover, and more importantly, we constructed a computable no-go conversion witness that is an improvement over the negativity, the previously best-known computable entanglement monotone, in determining non-convertibility of arbitrary states in quantum systems of any size. For certain states of two-qubit systems, this conversion witness is also better than both the negativity and the entanglement of formation.

The main goal of this paper was to illustrate the importance of conversion witnesses in the context of entanglement theory, but entanglement is not the only concept in quantum information that can be considered as a resource. Other ``resource theories'' that are determined by sets of restricted operations other than LOCC can be considered. The task of determining what is possible given certain allowable quantum operations is exactly the study of \emph{quantum resource theories} (for a recent review, see \cite{Horodecki2013}). Although our treatment of conversion witnesses was limited to entanglement, the concept of a computable resource conversion witness could (and should) be studied in all resource theories. Furthermore, beyond the quantum framework, resource theories can be studied as mathematical entities in their own right \cite{Fritz2013,Coecke2014}, and conversion witnesses may have applications in such abstract resource theories as well.

While we have only studied examples of conversion witnesses in the resource theory of entanglement, the search for conversion witnesses in other resource theories that yield improvements to the best-known monotones may be fruitful.

\appendix

\section{Proof of operator inequalities}
\label{app:opineqproofs}

We now prove the operator inequalities in \eqref{eq:posmatrices}. For positive semi-definite operators $A,B\in H_{n,+}$, we have
\begin{equation}
 (A-B)_+\leq A \hspace{5mm}\text{and}\hspace{5mm}(A-B)_-\leq B.
\end{equation}
Indeed, note that $(A-B)_+=P^\dagger(A-B)P$ where $P$ is a projector onto the subspace spanned by the eigenvectors of $A-B$ with corresponding positive eigenvalue. For any vector $\ket{\psi}\in\CC^n$, denote $\ket{\psi'}=P\ket{\psi}$. Since $A$ and $B$ are both positive operators, we have
\begin{align*}
 \bra{\psi}(A-B)_+\ket{\psi} 
 &=\bra{\psi'}(A-B)\ket{\psi'}\\
 &=\bra{\psi'}A\ket{\psi'} - \bra{\psi'}B\ket{\psi'}\\
 &\leq \bra{\psi'}A\ket{\psi'}\\
 &=\bra{\psi}P^\dagger A P \ket{\psi} \leq \bra{\psi}A\ket{\psi}
\end{align*}
since $P^\dagger A P\leq A$ for any projection operator $P$ and positive operator $A$. Thus $(A-B)_+\leq A$ as desired. 

The second inequality in \eqref{eq:posmatrices} follows from the first, since
\begin{align*}
 (A-B)_- &= (A-B)_+ - (A-B) \\&= B +\underbrace{(A-B)_+ - A}_{\leq 0} \leq B.
\end{align*}

\section{Computation of support function \texorpdfstring{$h_{\calN_c}(\tau)$}{hNc(tau)} and the witnesses \texorpdfstring{$W_{\tau}$}{Wtau} for Werner and isotropic~\texorpdfstring{$\tau$}{tau}}
In this section, we introduce the Werner states and isotropic states of a $d\times d$-system, then calculate $h_{\calN_c}(\tau)$ when $\tau$ is Werner or isotropic. This is used to determine the form of the witnesses $W_{\tau}$ when $\tau$ is Werner or isotropic, and new witnesses are defined by minimizing $W_{\tau}$ over the Werner and isotropic states.

\label{app:wernerisotropiccalculations}

\subsection{Werner states}
\label{app:Werner}
Recall that a Werner state is a state of a $d\times d$-system that is invariant under the action of all unitaries of the form $U\otimes U$. That is, a state $\rho$ is Werner if 
\[
 \rho=U\otimes U \rho U^\dagger \otimes U^\dagger
\]
for all unitaries $U$ acting on a $d$-dimensional Hilbert space.  The Werner states of a $d\times d$-system may be parametrized by a one-dimensional parameter as in \eqref{eq:wernerparameter}. It is important to determine the negative part of the partial transpose of the Werner states. First note that
\[
 (\omega_\alpha^d)^\Gamma = \frac{\alpha}{d}\ketbra{\Phi}{\Phi}+\frac{d-\alpha}{d(d^2-1)}\left(I-\ketbra{\Phi}{\Phi}\right).
\]
Thus the negativity of the Werner states is given by
\[
 N(\omega_\alpha^d)=\left\{\begin{array}{ll}
                      \frac{-\alpha}{d}, & -1\leq \alpha <0\\
                      0, & 0\leq\alpha\leq 1,
                      \end{array}\right.
\]
and the negative part of the partial transpose of $\omega_\alpha^d$ is
\[
 (\omega_\alpha^d)^\Gammaminus = N(\omega_\alpha^d) \ketbra{\Phi}{\Phi}.
\]
Note that the maximum negativity of all $d$-dimensional Werner states is $\frac{1}{d}$. 

We now make use of the high degree of symmetry of the Werner states to calculate the support function $h_{\calN_c}$ for Werner states. Recall that the Werner states are invariant under the twirling operator
\begin{equation}\label{eq:twirlingUU}
 \calT_{U\otimes U}(\rho)= \tfrac{1}{\int dU} \int U\otimes U \,\rho \, U^\dagger\otimes U^\dagger\, dU.
\end{equation}
For all Werner states $\tau=\omega_\alpha^d$ we have $\calT_{U\otimes U}(\tau)=\tau$. 
Furthermore, the twirling operator in \eqref{eq:twirlingUU} is self-adjoint with respect to the Hilbert-Schmidt inner product, i.e.\ $\Tr\left[\calT_{U\otimes U}(A)B\right]=\Tr\left[A\,\calT_{U\otimes U}(B)\right]$ for all hermitian operators $A$ and $B$. 

To calculate $h_{\calN_c}$, suppose that the operator $\gamma^\star\in\calN_c$ is optimal such that $\Tr[\gamma^\star\tau]=h_{\calN_c}(\tau) = \max_{\gamma\in\calN_c}\Tr[\gamma \tau]$. Then for $\tau=\omega_\alpha^d$ we have
\begin{align*}
 h_{\calN_c}(\omega_\alpha^d) = \Tr[\gamma^\star\,\omega_\alpha^d] = \Tr[\calT_{U\otimes U}(\gamma^\star)\,\omega_\alpha^d].
\end{align*}
Note that $\calT_{U\otimes U}(\gamma^\star)$ is also a Werner state and $\calT_{U\otimes U}$ is a PPT operation, so the negativity of $\calT_{U\otimes U}(\gamma^\star)$ must not be greater than $c$. Hence, to calculate $h_{\calN_c}(\tau)$ whenever $\tau=\omega_\alpha^d$ is a Werner state, it suffices to maximize over the Werner states with negativity bounded by $c$, and thus
\[
 h_{\calN_c}(\omega_\alpha^d)  = \max_{\substack{\omega^d_{\beta}\text{ Werner}\\N(\omega^d_\beta)\leq c}} \Tr[\omega^d_\beta\omega_\alpha^d].
\]
Note that $\Tr[\omega^d_\beta\omega_\alpha^d]= \frac{d(\alpha\beta+1)-\alpha-\beta}{d(d^2-1)}$. 

The Werner states of a $d\times d$-system have negativity of at most $\frac{1}{d}$, so this calculation needs to be considered in two cases. If $c\geq \frac{1}{d}$, then the maximum is taken over $\beta\in[-1,1]$. Otherwise, if $0\leq c<\frac{1}{d}$, the Werner state with negativity $N(\omega^d_\beta)=c$ has $\beta=-cd$, and the maximum is taken over $\beta\in[-cd,1]$. Since the function $\Tr[\omega^d_\beta\omega_\alpha^d]$ is linear in $\beta$, it suffices to check only the endpoints of the range of $\beta$. The support function for Werner states evaluates to
\begin{equation*}
 h_{\calN_c}(\omega_\alpha^d)=\left\{\begin{array}{ll}
                                 \displaystyle\frac{\alpha-d-cd(d\alpha-1)}{d(d^2-1)}, &
                                 \begin{array}{ll}
				  -1\leq \alpha < \frac{1}{d},\\ \hspace{2.8mm}0\leq c< \frac{1}{d}
                                 \end{array}\\
                                 \\
                                 \displaystyle\frac{1-\alpha}{d(d-1)}, & \begin{array}{ll}
				  -1\leq \alpha < \frac{1}{d},\\ \hspace{2.8mm}\frac{1}{d}\leq c
                                 \end{array}\\
                                 \\
                                 \displaystyle\frac{\alpha+1}{d(d+1)}, & \frac{1}{d}\leq \alpha\leq1.
                                \end{array}\right.
\end{equation*}

We are now ready to minimize the witnesses $W_{\tau}$ over all Werner states $\tau=\omega_\alpha^d$ to define the new witness
\begin{equation}\label{eq:wernerwitnessdefine}
W_{\text{Wer}}(\rho,\sigma):=d^2\,\min_{\omega_\alpha^d} W_{\omega_\alpha^d}(\rho,\sigma).
\end{equation}
The value of the witness $W_{\omega_\alpha^d}$ is piecewise linear in $\alpha$, so it suffices to check only the endpoints of the segments in which $W_{\omega_\alpha^d}$ is linear, i.e.\ $\alpha=-1$, $\alpha=\frac{1}{d}$, and  $\alpha=1$. For any $\alpha\geq\frac{1}{d}$, note that
\begin{align*}
W_{\omega_\alpha^d}(\rho,\sigma)\geq W_{\omega^d_{1/d}}(\rho,\sigma) = \frac{N(\rho)-N(\sigma)}{d^2} = \frac{1}{d^2}W_N(\rho,\sigma)
\end{align*}
for any states $\rho$ and $\sigma$. 
Performing the minimization in~\eqref{eq:wernerwitnessdefine}, this reduces to
\begin{equation}
 W_{\text{Wer}}(\rho,\sigma) = \min\left\{W_N(\rho,\sigma), \, \frac{2d}{d+1}W'_{\text{Wer}}(\rho,\sigma)\right\}
\end{equation}
where $W'_{\text{Wer}}$ is the sub-witness defined by
\[
 W'_{\text{Wer}}(\rho,\sigma) = \left\{\begin{array}{ll}
                                         \frac{dN(\tilde\rho)+1}{2}N(\rho) - \Tr[F_-\sigma^\Gammaminus], &  N(\tilde\rho) < \frac{1}{d}\\\\
                                         N(\rho)-\Tr[F_-\sigma^\Gammaminus], &  N(\tilde\rho) \geq \frac{1}{d}.
                                        \end{array}\right.
\]
Note that $\Tr[F_-\sigma^\Gammaminus]\leq \Tr[\sigma^\Gammaminus]=N(\sigma)$, where $F_-$ is the negative component of the flip operator. If $N(\sigma)\leq N(\rho)$ then the value of the witness $W'_\text{Wer}(\rho,\sigma)$ can be negative only when $\tfrac{dN(\tilde\rho)+1}{2}<1$. Hence, the new witness $W_{\text{Wer}}$ that is obtained by minimizing over the Werner states is only an improvement to the negativity if $N(\tilde\rho)<\frac{1}{d}$. So we may instead assume the simpler form
\begin{align}                                                                                                                                                                                                                                                                                                                                                                                                                                                         
      W'_{\text{Wer}}(\rho,\sigma) &=     \frac{dN(\tilde\rho)+1}{2}N(\rho) - \Tr[F_-\sigma^\Gammaminus]\\
      &=     \frac{d\Tr[\rho^{\Gammaminus\Gammaminus}]+\Tr[\rho^\Gammaminus]}{2} - \Tr[F_-\sigma^\Gammaminus]\nonumber
\end{align}
for the witness.

\subsection{Isotropic states}
\label{app:supportforisotropic}

Recall that the isotropic states~\cite{Horodecki1999} of a $d\times d$-system are those that are invariant under the action of all unitaries of the form $U\otimes \bar{U}$, where $\bar{U}$ denotes the complex conjugate of $U$. The Werner states of a $d\times d$-system may be parametrized by a one-dimensional parameter as in \eqref{eq:wernerparameter}. As in the previous section, it is useful to determine the negative part of the partial transpose of isotropic states. The flip operator can be written as
\begin{equation}\label{eq:flipdefine}
 F=\underbrace{\sum_{i=1}^d\ketbra{ii}{ii} + \sum_{i<j}\ketbra{\psi_{ij}^+}{\psi_{ij}^+}}_{F_+} - \underbrace{\sum_{i<j}\ketbra{\psi_{ij}^-}{\psi_{ij}^-}}_{F_-},
\end{equation}
where $\ket{\psi_{ij}^\pm}=\frac{1}{\sqrt{2}}\left(\ket{ij}\pm\ket{ji}\right)$, and we can split $F$ into its positive and negative components $F=F_+-F_-$. Note that
\[
 (\eta_\beta^d)^\Gamma = \frac{1-d\beta}{d(d-1)}F_- + \frac{1+d\beta}{d(d+1)}F_+
\]
and that the negativity $N(\eta_\beta^d))=\Tr[(\eta_\beta^d)^\Gammaminus]$ of the isotropic states can be given by
\[
 N(\eta_\beta^d)=\left\{\begin{array}{cc}
                      \frac{d\beta-1}{2} ,& \frac{1}{d}< \beta\leq 1\\
                       0, & 0\leq \beta\leq \frac{1}{d}.
                      \end{array}\right.
\]
Furthermore, note that $(\eta_\beta^d)^\Gammaminus = N(\eta_\beta^d) \frac{2}{d(d-1)}F_-$. 
 
As before, we make use of the high degree of symmetry to calculate the value of the support function $h_{\calN_c}$ for isotropic states. Recall that the isotropic states are invariant under the twirling operator
\begin{equation}\label{eq:twirlingUUbar}
 \calT_{U\otimes \bar{U}}(\rho)= \tfrac{1}{\int dU} \int U\otimes \bar{U} \,\rho \, U^\dagger\otimes \bar{U}^\dagger\, dU.
\end{equation}
For isotropic states $\tau=\eta_\beta^d$ we have $\calT_{U\otimes\bar{U}}(\tau)=\tau$. Furthermore, the twirling operator in \eqref{eq:twirlingUUbar} is also self-adjoint with respect to the Hilbert-Schmidt inner product.

Using the same arguments as for Werner states, it suffices to optimize only over the isotropic states when calculating the support the function $h_{\calN_c}$ for isotropic states. So the support function reduces to 
\[
 h_{\calN_c}(\eta_\beta^d)  = \max_{\substack{\eta^d_\alpha \text{ isotropic}\\N(\eta^d_\alpha)\leq c}} \Tr[\eta^d_\alpha\eta_\beta^d].
\]
Note that  $\Tr[\eta^d_\alpha\eta_\beta^d]=\frac{d^2\alpha\beta-\alpha-\beta+1}{d^2-1}$.

The isotropic states of a $d\times d$-system have negativity of at most $\frac{d-1}{2}$, so this calculation must be considered in two cases. If $c\geq \frac{d-1}{2}$, then the maximum is taken over $\alpha\in[0,1]$. Otherwise, if $0\leq c<\frac{d-1}{2}$, the isotropic state with negativity $N(\eta^d_\alpha)=c$ has $\alpha=\frac{2c+1}{d}$, and the maximum is taken over $\alpha\in[0,\frac{2c+1}{d}]$. Since the function $\Tr[\eta^d_\alpha\eta_\beta^d]$ is linear in $\alpha$, it suffices to check only the endpoints of the range of $\alpha$. The support function for isotropic states evaluates to
\begin{widetext}
\begin{equation*}
 h_{\calN_c}(\eta_\beta^d) = \left\{ \begin{array}{ll}
                                    \beta+\displaystyle\frac{(1+2c-d)(d^2\beta-1)}{d(d^2-1)},    &  \frac{1}{d^2}< \beta \leq 1, \, 0\leq c <\frac{d-1}{2}\\\\
                                    \beta,    &  \frac{1}{d^2}< \beta \leq 1, \, \frac{d-1}{2}\leq c \\\\
                                    \displaystyle\frac{1-\beta}{d^2-1},    &  0\leq \beta \leq \frac{1}{d^2}.
                                   \end{array} \right.
\end{equation*}
\end{widetext}

The calculations for minimizing the witnesses $W_{\tau}$ over all isotropic states $\tau=\eta_\beta^d$ to define the new witness
\begin{equation}\label{eq:isotropicwitnessdefine}
W_{\text{iso}}(\rho,\sigma):=d^2\,\min_{\eta_\beta^d} W_{\eta_\beta^d}(\rho,\sigma)
\end{equation}
are analogous to those for the Werner states. Performing the minimization in~\eqref{eq:isotropicwitnessdefine}, this reduces to
\begin{equation}
 W_{\text{iso}}(\rho,\sigma) = \min\left\{W_N(\rho,\sigma), \, W'_{\text{iso}}(\rho,\sigma)\right\}
\end{equation}
where $W'_{\text{iso}}$ is the sub-witness defined by
\[
 W'_{\text{iso}}(\rho,\sigma) = \left\{\begin{array}{ll}
                                         \frac{2N(\tilde\rho)+1}{d}N(\rho) - \bra{\Phi}\sigma^\Gammaminus\ket{\Phi}, &  N(\tilde\rho) < \frac{d-1}{2}\\\\
                                         N(\rho)-\bra{\Phi}\sigma^\Gammaminus\ket{\Phi}, &  N(\tilde\rho) \geq \frac{d-1}{2}.
                                        \end{array}\right.
\]
Note that $\bra{\Phi}\sigma^\Gammaminus\ket{\Phi}\leq \Tr[\sigma^\Gammaminus]=N(\sigma)$. If $N(\sigma)\leq N(\rho)$ then the value of the witness $W'_\text{iso}(\rho,\sigma)$ can be negative only when $\tfrac{2N(\tilde\rho)+1}{d}<1$. Hence, the new witness $W_{\text{iso}}$ that is obtained by minimizing over the Werner states is only an improvement to the negativity if $N(\tilde\rho)<\frac{d-1}{2}$. So we may instead assume the simpler form
\begin{align*}                                                                                                                                                                                                                                                                                                                                                                                                                                                         
      W'_{\text{iso}}(\rho,\sigma) &=     \frac{2N(\tilde\rho)+1}{d}N(\rho) - \bra{\Phi}\sigma^\Gammaminus\ket{\Phi}\\
      &=\frac{2\Tr[\rho^{\Gammaminus\Gammaminus}]+\Tr[\rho^\Gammaminus]}{d}-\Tr[\ketbra{\Phi}{\Phi}\sigma^\Gammaminus]\nonumber
\end{align*}
for the witness.

\section{Further no-go witnesses for PPT operations}
\label{app:further}

The key idea in this paper is that we can construct a no-go witness by determining if a state $\sigma$ is outside of the PPT-orbit of the initial state $\rho$. However, the PPT-orbit is difficult to characterize, so we resort to other more easily computable techniques. Namely, instead of considering the orbits $\calC_\Gamma(\tilde\rho)$, we consider the larger sets $\calN_{N(\tilde\rho)}$ of states that have negativity less $N(\tilde{\rho})$. In this appendix, we construct further sets of states that contain the orbit $\calC_\Gamma(\tilde\rho)$ but are smaller than~$\calN_{N(\tilde\rho)}$.  

Define the following set of positive hermitian matrices:
\begin{equation*}
 \calN_1(\rho)=\left\{\gamma\in H_{n,+}\,\middle|\,\Tr[\gamma]=N(\rho),\, N(\gamma)\leq N(\rho^\Gammaminus\!)\right\}.
\end{equation*}
Note that we have the following containment:
\begin{equation*}
 \calC_\Gamma(\rho^\Gammaminus\!)\subseteq \calN_1(\rho).
\end{equation*}
Indeed, for every $\gamma\in\calC_\Gamma(\rho^\Gammaminus\!)$, we have $\gamma=\calE(\rho^\Gammaminus\!)$ for some operation $\calE\in\calC_\Gamma$, and thus $N(\gamma)=N\bigl(\calE(\rho^\Gammaminus\!)\bigr)\leq N(\rho^\Gammaminus\!)$ since the negativity is a monotone under PPT-operations. Furthermore, since each $\calE\in\calC_\Gamma$ is trace-preserving, we have $\Tr[\gamma]=\Tr[\calE(\rho^\Gammaminus\!)]=\Tr[\rho^\Gammaminus\!] =N(\rho)$.

Each $\gamma\in\calC_\Gamma(\rho^\Gammaminus\!)$ is given by $\gamma=\calE(\rho^\Gammaminus\!)$ for some PPT-operation. For any positive operator $A$ and PPT operation $\Lambda\in\calC_\Gamma$, we recall the important operator inequality from section III
\[
 A^\Gammaminus \leq \Lambda^\Gamma(A^\Gammaminus).
\]
We can use this inequality by setting $A=\rho^\Gammaminus$ which gives us
\begin{equation}\label{eq:gammagammaminus}
 \gamma^\Gammaminus \leq \calE(\rho^{\Gammaminus\Gammaminus}\!),
\end{equation}
where $\rho^{\Gammaminus\Gammaminus}=(\rho^\Gammaminus\!)^\Gammaminus$.
Taking the trace of both sides in \eqref{eq:gammagammaminus} yields $N(\gamma)\leq N(\rho^\Gammaminus\!)$, but much of the information in~\eqref{eq:gammagammaminus} is lost. Note that $\Tr\bigl[\calE(\rho^{\Gammaminus\Gammaminus})\bigr]=N(\rho^\Gammaminus\!)$, and thus the relation in \eqref{eq:gammagammaminus} implies that there exists an $\eta\in H_{n,+}$ with $\Tr[\eta]=N(\rho^\Gammaminus\!)$ and $N(\eta)\leq N(\rho^{\Gammaminus\Gammaminus})$ such that $\gamma^\Gammaminus\leq \eta$. In particular, we can choose $\eta=\calE(\rho^{\Gammaminus\Gammaminus})$. So we can make a refinement of $\calN_1(\rho)$ by defining the set
\begin{widetext}
\begin{equation*}
 \calN_2(\rho)=\left\{\gamma\in H_{n,+}\,\middle|\,\Tr[\gamma]=N(\rho),\, \exists\eta\in H_{n,+} \text{ with } \Tr[\eta]=N(\rho^{\Gammaminus})\text{ and }N(\eta)\leq N(\rho^{\Gammaminus\Gammaminus}\!)\text{ such that } \gamma^\Gammaminus\leq \eta\right\}.
\end{equation*}
By construction we have that $\calC_\Gamma(\rho^\Gammaminus\!)\subseteq \calN_2(\rho)$. Furthermore, for $\gamma\in\calN_2(\rho)$, we have $N(\gamma)=\Tr[\gamma^\Gammaminus\!]\leq \Tr[\eta]=N(\rho^\Gammaminus\!)$ and thus $\gamma\in\calN_1(\rho)$. So we have the following containment structure:
\[
 \calC_\Gamma(\rho^\Gammaminus\!)\subseteq\calN_2(\rho)\subseteq \calN_1(\rho).
\]
For $\gamma=\calE(\rho^\Gammaminus)$ we can always choose $\eta=\calE(\rho^{\Gammaminus\Gammaminus}\!)$, so we have $\eta^\Gammaminus\leq\calE(\rho^{\Gammaminus\Gammaminus\Gammaminus})$. However, as before, much information is ``lost'' by taking the trace of this and distilling it down to $N(\eta)\leq N(\rho^{\Gammaminus\Gammaminus})$. 

Now let $\eta_1=\eta$. Continuing in a manner analogous to \eqref{eq:gammagammaminus}, we see that 
\begin{equation*}
 \eta_1^\Gammaminus\leq \calE(\rho^{\Gammaminus\Gammaminus\Gammaminus}\!)
\end{equation*}
implies that there exists a $\eta_2\in H_{n,+}$ with $\Tr[\eta_2]=\Tr[\calE(\rho^{\Gammaminus\Gammaminus\Gammaminus}\!)]=N(\rho^{\Gammaminus\Gammaminus}\!)$ and $N(\eta_2)\leq N(\rho^{\Gammaminus\Gammaminus\Gammaminus}\!)$ such that $\eta_1^\Gammaminus\leq \eta_2$. In particular, as before, we can choose $\eta_2=\calE(\rho^{\Gammaminus\Gammaminus\Gammaminus})$. So we can define the set
\begin{equation*}
 \calN_3(\rho)=\left\{\gamma\in H_{n,+}\,\middle|\,
 \begin{array}{c}
  \Tr[\gamma]=N(\rho),\, \exists\eta_1,\eta_2\in H_{n,+} \text{ such that }\\ \Tr[\eta_1]=N(\rho^{\Gammaminus})\text{ and }\gamma^\Gammaminus\leq \eta_1 ,\\ \Tr[\eta_2]=N(\rho^{\Gammaminus\Gammaminus})\text{ and }\eta_1^\Gammaminus\leq \eta_2,\\
  \text{and }N(\eta_2)\leq N(\rho^{\Gammaminus\Gammaminus\Gammaminus})
 \end{array}
 \right\},
\end{equation*}
such that $\calC_\Gamma(\rho^\Gammaminus\!)\subseteq\calN_3(\rho)\subseteq\calN_2(\rho)\subseteq \calN_1(\rho).$ Carrying this process out, for each $k$ we can define the set
\begin{equation*}
 \calN_k(\rho)=\left\{\gamma\in H_{n,+}\,\middle|\,
 \begin{array}{c}
  \Tr[\gamma]=N(\rho),\, \exists\eta_1,\dots,\eta_{k-1}\in H_{n,+} \text{ such that}\\  \gamma^{\Gamma_{\!-}^{(j)}}\leq \eta_j\text{, }\Tr[\eta_j]=N(\rho^{\Gamma_{\!-}^{(j)}})\text{, and }N(\eta_j)\leq N(\rho^{\Gamma_{\!-}^{(j+1)}})\\
  \forall j\in\{1,\dots,k-1\}
 \end{array}
 \right\},
\end{equation*}
where $\rho^{\Gamma_{\!-}^{(j)}}=\rho^{\scriptstyle\overbrace{\scriptstyle\Gammaminus\cdots\Gammaminus}^{j\text{ times}}}$. This gives us the containments
\[
 \calC_\Gamma(\rho^\Gammaminus\!)\subseteq\cdots\subseteq\calN_3(\rho)\subseteq\calN_2(\rho)\subseteq \calN_1(\rho).
\]

\end{widetext}

\section{Generalized Werner states}
\label{app:rhoabstates}

In this section, we consider the generalized Werner states, first analyzed in the context of the distillability problem~\cite{DiVincenzo2000}. This is a symmetric class of states that contains the Werner states. The support function $h_{\calN_c}(\tau)$ can be computed when $\tau$ is a generalized Werner state, which we show here, so a conversion witness $W_\text{Gwer}$ can be constructed by minimizing over all $W_\tau(\rho,\sigma)$ where $\tau$ is a generalized Werner state. However, in the following computation, we show that $W_\text{wer}\succeq W_\text{Gwer}$, so this new witness is not any better than $W_\text{wer}$.

The generalized Werner states of a bipartite $d\times d$-system are defined in the following manner. Consider the subgroup of $d$-dimensional unitary matrices that are diagonal with respect to some fixed basis $\{\ket{i}\}$. These are matrices $U$ such that $\bra{j}U\ket{i}=\delta_{ij}e^{i\theta_i}$. A state $\rho$ is a generalized Werner state if it is invariant under any bipartite unitary $U\otimes U$, where $U$ is diagonal.  These states may be parametrized by two real parameters. For $a,b\in[0,1]$ with $a+b\leq 1$, the generalized Werner states can be given by
\begin{multline*}
 \rho_{ab}=(1-a-b)\frac{1}{d}\overbrace{\sum_{i=0}^{d-1}\ketbra{ii}{ii}}^{P} +a \frac{2}{d(d-1)} \overbrace{\sum_{i<j}\ketbra{\psi^-_{ij}}{\psi^-_{ij}}}^{P_-}\\+b \frac{2}{d(d-1)} \underbrace{\sum_{i<j}\ketbra{\psi^+_{ij}}{\psi^+_{ij}}}_{P_+}.
\end{multline*}
For brevity, we can instead write these states as
\[
 \rho_{ab}=(1-a-b)\tfrac{1}{d}P+ a\tfrac{2}{d(d-1)}P_- + b\tfrac{2}{d(d-1)}P_+
\]
where $P$, $P_+$, and $P_-$ are orthogonal projection matrices with $P+P_++P_-= I$. 

\begin{figure}[t!]\centering
\begin{tikzpicture}[scale=0.8, every node/.style={scale=0.8}]
  \coordinate[label = {below left:$0$}] (O) at (0,0);
  \coordinate[label = {$1$}] (1a) at (8,-.55);
  \coordinate[label = {left:$1$}] (1b) at (0,8);
  \coordinate[label = {left:$\tfrac{d-1}{d}$}] (NPTedge) at (0,7);
  \coordinate[label = {left:$\tfrac{d-1}{d+1}$}] (Werner) at (0,6);
  \coordinate[label = {$\tfrac{1}{2}$}] (ahalf) at (4,-.65);
  \coordinate[label = {left:$\tfrac{1}{2}$}] (bhalf) at (0,4);
  \draw[->] (0,0) -- (9,0) coordinate[label = {below:$a$}] (amax);;
  \draw[->] (0,0) -- (0,9) coordinate[label = {left:$b$}] (bmax);
  \draw[] (0,8) -- (8,0);
  \draw[gray,dashed,opacity=.5] (0,4) -- (4,4);
  
  \draw[gray]  (0,6) -- (8,0);
  
  \draw (0,4) -- (-.1,4);
  \draw (0,6) -- (-.1,6);
  \draw (0,7) -- (-.1,7);
  \draw (0,8) -- (-.1,8);
  \draw (4,0) -- (4,-.1);
  \draw (8,0) -- (8,-.1);

  \draw[fill=blue,opacity=.4]  (0,7) -- (0,8) -- (4,4) -- cycle;
  \draw[fill=blue,opacity=.4]  (4,0) -- (4,4) -- (8,0) -- cycle;
  
  \node[anchor=east] at (6,6,0) (NPT) {NPT};
  \draw (NPT) edge[out=160,in=45,->] (1.5,6);
  \draw (NPT) edge[out=300,in=45,->] (5,2.75);
  
  \node at (2,2) (PPT) {PPT};
  
  \node[anchor=west] at (7,2) (Wer) {Werner states};
  \draw (Wer) edge[out=180,in=45,->] (6,1.5);
\end{tikzpicture}
\caption{Schematic of the $\rho_{ab}$ states. The PPT region is shown in white and the two NPT regions are shaded in purple. The one-dimensional family of states with $b=\frac{d-1}{d+1}(1-a)$ are the well-known Werner states.}
\label{fig:rhoab}
\end{figure}
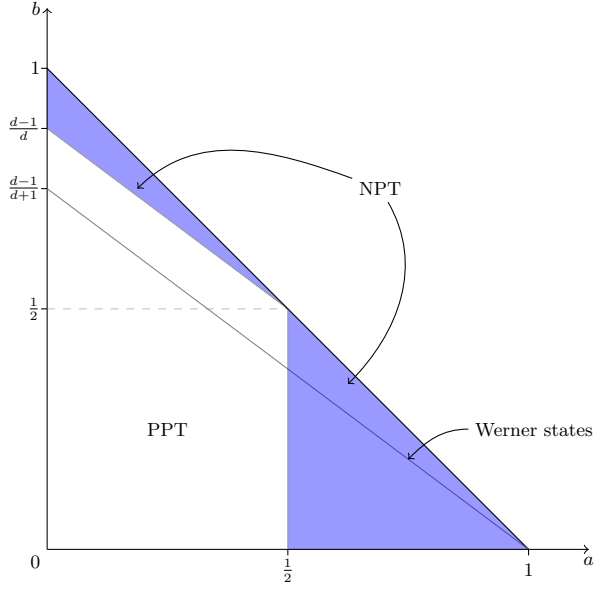

 To determine the negativity of the $\rho_{ab}$ states, we need to compute their partial transpose.  This can be given by
\begin{align*}
\rho_{ab}^\Gamma 
&= \frac{1-a-b}{d}\sum_{i=0}^{d-1}\ketbra{ii}{ii}  + \frac{b-a}{d(d-1)}\sum_{i\neq j}\ketbra{ii}{jj} \\
&\hspace{15mm}+ \frac{b+a}{d(d-1)}\sum_{i\neq j}\ketbra{ij}{ij}\\
&=\left(\frac{1-a-b}{d}+\frac{a-b}{d(d-1)}\right)P  +\frac{b-a}{d(d-1)}\overbrace{\sum_{i,j}\ketbra{ii}{jj}}^{d\ketbra{\Phi}{\Phi}}\\
&\hspace{15mm}+ \frac{b+a}{d(d-1)}\sum_{i\neq j}\ketbra{ij}{ij}.
\end{align*}
Note that $\ketbra{\Phi}{\Phi}$ is another projection operator with $\ketbra{\Phi}{\Phi}< P$, and thus $P'=P- \ketbra{\Phi}{\Phi}$ is also a projection operator. Hence
\begin{align*}
 \rho_{ab}^\Gamma 
 &= \left(\frac{1-a-b}{d}+\frac{a-b}{d(d-1)}\right)P' + \frac{b+a}{d(d-1)}\sum_{i\neq j}\ketbra{ij}{ij} \\
 &\hspace{15mm}+ \underbrace{\left(\frac{1-a-b}{d}+\frac{a-b}{d(d-1)} + \frac{b-a}{d-1}\right)}_{\frac{1-2a}{d}}\ketbra{\Phi}{\Phi} \\
 &= \frac{a(d-2)+bd-d+1}{d(d-1)}P'  + \frac{b+a}{d(d-1)}\sum_{i\neq j}\ketbra{ij}{ij} \\
 &\hspace{15mm}+\frac{1-2a}{d}\ketbra{\Phi}{\Phi}.
\end{align*}
Note that only the coefficients in front of the $P'$ and $\ketbra{\Phi}{\Phi}$ can be negative. In particular, the negativity of $\rho_{ab}$ is given by
\[
 N(\rho_{ab})=\left\{\begin{array}{ll}
                      \frac{2a-1}{d}, & \frac{1}{2}<a\leq 1-b\\
                      \frac{a(d-2)+bd-d+1}{d}, & \frac{d-1-a(d-2)}{d}<b\leq 1-a\\
                      0,& \text{else}.
                     \end{array}
\right.
\]
A schematic of the $\rho_{ab}$ states can be seen in FIG.\ \ref{fig:rhoab}. 

\begin{figure}[t]
 \centering
 \begin{tikzpicture}[scale=0.8, every node/.style={scale=0.8}]
  \coordinate[label = {below left:$0$}] (O) at (0,0);
  \coordinate[label = {$1$}] (1a) at (8,-.55);
  \coordinate[label = {left:$1$}] (1b) at (0,8);
  \coordinate[label = {left:$\tfrac{d-1}{d}$}] (NPTedge) at (0,7);
  \coordinate[label = {$\tfrac{1}{2}$}] (ahalf) at (4,-.65);
  \coordinate[label = {left:$\tfrac{1}{2}$}] (bhalf) at (0,4);  
  \coordinate[label = {left:$\tfrac{1-cd}{2}$}] () at (0,2);
  \coordinate[label = {left:$\tfrac{1+cd}{2}$}] () at (0,6);
  \coordinate[label = {below:$\tfrac{1+cd}{2}$}] () at (6,0);
  \coordinate[label = {below:$\tfrac{1-cd}{2}$}] () at (2,0);
  \coordinate[label = {left:$\tfrac{d(c+1)-1}{d}$}] () at (-.5,7.5);
  
  \draw[->] (0,0) -- (9,0) coordinate[label = {below:$a$}] (amax);;
  \draw[->] (0,0) -- (0,9) coordinate[label = {left:$b$}] (bmax);
  \draw[] (0,7.5) -- (2,6) -- (6,2) -- (6,0);
  \draw[gray,opacity=.5] (0,8) -- (8,0);
  \draw[gray,dashed,opacity=.5] (0,6) -- (2,6);
  \draw[gray,dashed,opacity=.5] (0,2) -- (6,2);
  \draw[gray,dashed,opacity=.5] (2,0) -- (2,6);  
  
  \draw[fill=red,opacity=.4]  (0,7) -- (0,7.5) -- (2,6) -- (4,4) -- cycle;
  \draw[fill=red,opacity=.4]  (4,0) -- (4,4) -- (6,2) -- (6,0)-- cycle;
  \draw[fill=red,opacity=.4] (0,0) -- (0,7.5) -- (2,6) -- (6,2) -- (6,0)-- cycle;
  
  \draw (0,4) -- (-.1,4);
  \draw (0,7) -- (-.1,7);
  \draw (0,8) -- (-.1,8);
  \draw (4,0) -- (4,-.1);
  \draw (8,0) -- (8,-.1);  
  \draw (6,0) -- (6,-.1);
  \draw (0,6) -- (-.1,6);
  \draw (2,0) -- (2,-.1);
  \draw (0,2) -- (-.1,2);
  \draw (0,7.5) -- (-.5,7.5);
\end{tikzpicture}
  \caption{The negativity sets $\calN_c$ restricted to the $\rho_{ab}$ states. For a $c\geq 0$, the set of $\rho_{ab}$ states with negativity $N(\rho_{ab})\leq c$ is shaded dark red. The PPT states with negativity $N(\rho_{ab})=0$ are contained in $\calN_c$ for all $c\geq0$, and are shaded light red.}
 \label{fig:rhoabNc}
\end{figure}
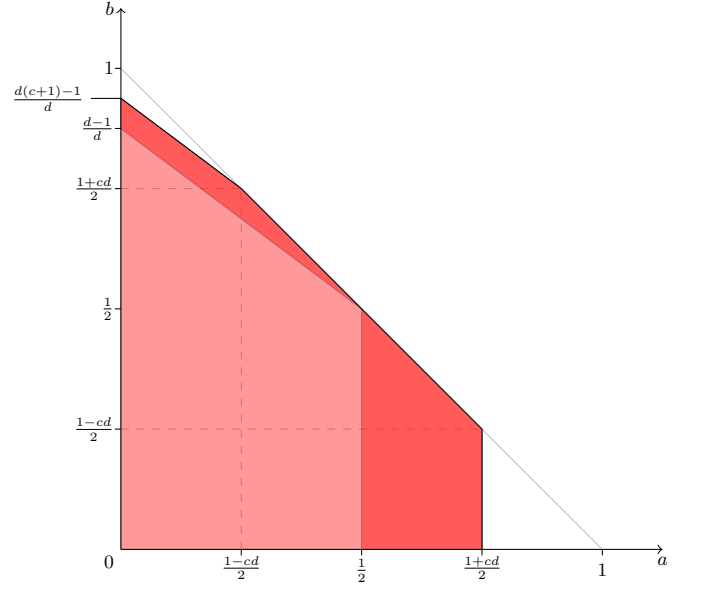

For a fixed state $\rho_{ab}$ and a value $0\leq c$, we want to determine the value of the support function $h_{\calN_c}(\rho_{ab})$. Due to the symmetry of the generalized Werner states, this is reduced to determining $\max\left\{\Tr[\rho_{ab}\rho_{a'b'}]\,\middle|\,N(\rho_{a',b'})\leq c\right\}$. For different values of~$c$, the sets of states $\rho_{a'b'}$ with negativity not greater than~$c$ are displayed in FIG.\ \ref{fig:rhoabNc}.

We now determine $\Tr[\rho_{ab}\rho_{a'b'}]$ for a fixed $\rho_{ab}$:
\begin{align}
\Tr&[\rho_{ab}\rho_{a'b'}]= \nonumber\\
&= \frac{(1-a-b)(1-a'-b')}{d} + \frac{2}{d(d-1)}(aa'+bb')\nonumber\\
&=a'\left(\frac{\frac{d+1}{d-1}a+b-1}{d}\right) + b'\left(\frac{\frac{d+1}{d-1}b+a-1}{d}\right) \nonumber\\
&\hspace{15mm}+ \frac{1-a-b}{d}\label{eq:appTrara'b'}.
\end{align}
Note that this is a linear function of $a'$ and $b'$. Let 
\[
A=\frac{d+1}{d-1}a+b-1\hspace{5mm}\text{and}\hspace{5mm}B=\frac{d+1}{d-1}b+a-1. 
\]
To maximize the quantity in \eqref{eq:appTrara'b'}, it suffices to find $a'$ and $b'$ that maximize the function $f(a',b')=a'A+b'B$. Since $a'$ and $b'$ are non negative, we should split the problem into cases when $A$ and $B$ are positive and negative. These regions are shown in FIG.\ \ref{fig:rhoabABregions}. If $A$ is negative, then we should choose $a'$ to be zero to maximize $f(a',b')$, whereas if $A>0$ then $a'$ should be positive. An analogous statement can be made for $B$ and $b'$.
\begin{figure}
\centering
\begin{tikzpicture}[scale=0.8, every node/.style={scale=0.8}]
  \coordinate[label = {below left:$0$}] (O) at (0,0);
  \coordinate[label = {$1$}] (1a) at (8,-.55);
  \coordinate[label = {left:$1$}] (1b) at (0,8);
  \coordinate[label = {left:$\tfrac{d-1}{d+1}$}] () at (0,6);
  \coordinate[label = {below:$\tfrac{d-1}{d+1}$}] () at (6,0);
  \coordinate[label = {left:$\tfrac{d-1}{2d}$}] () at (0,3.4375);
  \coordinate[label = {below:$\tfrac{d-1}{2d}$}] () at (3.4375,0);
  \coordinate[label = {$\tfrac{1}{2}$}] (ahalf) at (4,-.65);
  \coordinate[label = {left:$\tfrac{1}{2}$}] (bhalf) at (0,4);
  \draw[->] (0,0) -- (9,0) coordinate[label = {below:$a$}] (amax);;
  \draw[->] (0,0) -- (0,9) coordinate[label = {left:$b$}] (bmax);

  \draw[gray,dashed,opacity=.5]  (3.4375,0) -- (3.4375,3.4375);
  \draw[gray,dashed,opacity=.5]  (0,3.4375) -- (3.4375,3.4375);
  \draw[gray,dashed,opacity=.5]  (4,0) -- (4,4);
  \draw[gray,dashed,opacity=.5]  (0,4) -- (4,4);
  
  \draw[gray,opacity=.5] (0,8) -- (8,0);
  \draw[fill=brown,opacity=.4] (8,0) -- (4,4)--(3.4375,3.4375)-- cycle;
  \draw[fill=green,opacity=.4] (0,8) -- (4,4)--(3.4375,3.4375)-- cycle;
  \draw[fill=gray,opacity=.8] (0,8) -- (4,4)--(3.4375,3.4375)-- cycle;
  \draw[fill=blue,opacity=.4] (0,0) -- (0,6) -- (8,0)-- cycle;
  \draw[fill=red,opacity=.4] (0,0) -- (6,0) -- (0,8)-- cycle;
  \draw[fill=gray,opacity=.4] (8,0) -- (3.4375,3.4375) -- (0,8)-- cycle;
  
  \node at (2,2) (ABneg) {$A,B<0$};
  \node at (6.4,.5) (AnegBpos) {$\begin{array}{cc} A>0\\B<0 \end{array}$};
  \node at (.5,6.4) (AposBneg) {$\begin{array}{cc} A<0\\B>0 \end{array}$};
  \node[anchor=west] at (3.5,5) (0<A<B) {$0<A\leq B$};
  \node[anchor=west] at (4.5,4) (0<B<A) {$0<B\leq A$};
  
  \draw (0<A<B) edge[out=225,in=25,->] (3.5,4.0);
  \draw (0<B<A) edge[out=270,in=65,->] (4.6,3.0);
    
  \draw (0,4) -- (-.1,4);
  \draw (0,8) -- (-.1,8);
  \draw (0,6) -- (-.1,6);
  \draw (6,0) -- (6,-.1);
  \draw (4,0) -- (4,-.1);
  \draw (8,0) -- (8,-.1);
  \draw (3.4375,0) -- (3.4375,-.1);
  \draw (0,3.4375) -- (-.1,3.4375);
\end{tikzpicture}
 \caption{The region where $A<0$ and $B<0$ is purple. The region where $A<0$ but $B>0$ is blue while the region where $A>0$ but $B<0$ is red. The region where both $A>0$ and $B>0$ is split into two further regions. The region with $A\leq B$ is green, while the region with $B\leq A$ is brown.}
 \label{fig:rhoabABregions}
\end{figure}
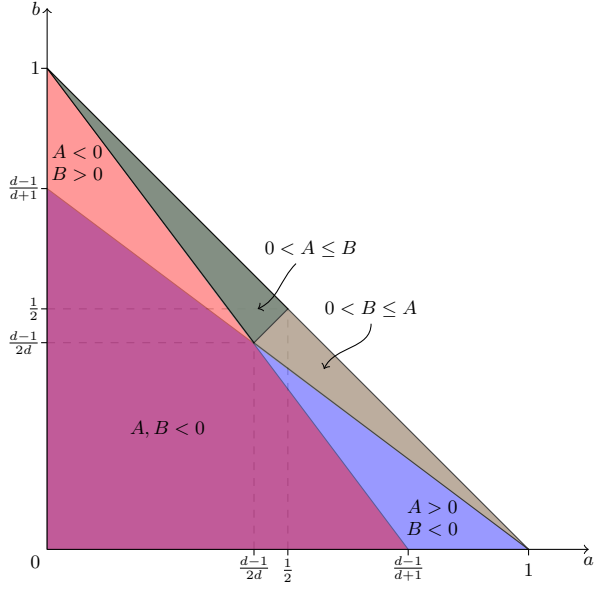

\begin{figure}[t!]
\centering
\begin{tikzpicture}[scale=0.8, every node/.style={scale=0.8},dot/.style = {
      draw,
      fill = black,
      circle,
      inner sep = 0pt,
      minimum size = 4pt}]
  \coordinate[label = {below left:$0$}] (O) at (0,0);
  \coordinate[label = {$1$}] (1a) at (8,-.55);
  \coordinate[label = {left:$1$}] (1b) at (0,8);
  \coordinate[label = {left:$\tfrac{d-1}{d+1}$}] () at (0,6);
  \coordinate[label = {below:$\tfrac{d-1}{d+1}$}] () at (6,0);
  \coordinate[label = {left:$\tfrac{d-1}{2d}$}] () at (0,3.4375);
  \coordinate[label = {below:$\tfrac{d-1}{2d}$}] () at (3.4375,0);
  \coordinate[label = {$\tfrac{1}{2}$}] (ahalf) at (4,-.65);
  \coordinate[label = {left:$\tfrac{1}{2}$}] (bhalf) at (0,4);
  \draw[->] (0,0) -- (9,0) coordinate[label = {below:$a$}] (amax);;
  \draw[->] (0,0) -- (0,9) coordinate[label = {left:$b$}] (bmax);

  \draw[gray,dashed,opacity=.5]  (3.4375,0) -- (3.4375,3.4375);
  \draw[gray,dashed,opacity=.5]  (0,3.4375) -- (3.4375,3.4375);
  \draw[gray,dashed,opacity=.5]  (4,0) -- (4,4);
  \draw[gray,dashed,opacity=.5]  (0,4) -- (4,4);
  
  \draw[gray] (0,8) -- (8,0);
  \draw[gray] (0,6) -- (8,0);
  \draw[gray] (0,8) -- (6,0);
  \draw[gray,thick] (0,8) -- (0,0) -- (8,0);
  \draw[gray] (3.4375,3.4375) -- (4,4);
  
  \node[dot,label = {above right:$G$}] at (0,0) () {};
  \node[dot,label = {      right:$H$}] at (0,6) () {};
  \node[dot,label = {above right:$J$}] at (0,8) () {};
  \node[dot,label = {above right:$K$}] at (6,0) () {};
  \node[dot,label = {above right:$L$}] at (8,0) () {};
  \node[dot,label = {below left :$M$}] at (3.4375,3.4375) () {};
  \node[dot,label = {above right:$N$}] at (4,4) () {};
      
  \draw (0,4) -- (-.1,4);
  \draw (0,8) -- (-.1,8);
  \draw (0,6) -- (-.1,6);
  \draw (6,0) -- (6,-.1);
  \draw (4,0) -- (4,-.1);
  \draw (8,0) -- (8,-.1);
  \draw (3.4375,0) -- (3.4375,-.1);
  \draw (0,3.4375) -- (-.1,3.4375);

\end{tikzpicture}
 \caption{The value of the witness $W_{\rho_{ab}}(\rho,\sigma)$ is linear in each region of FIG.\ \ref{fig:rhoabABregions}, so the optimal choice of $\rho_{ab}$ that minimizes the witness $W_{\rho_{ab}}(\rho,\sigma)$ must be at one of the vertices  $G$, $H$, $J$, $K$, $L$, $M$ or $N$.}
 \label{fig:rhoabABpoints}
\end{figure}
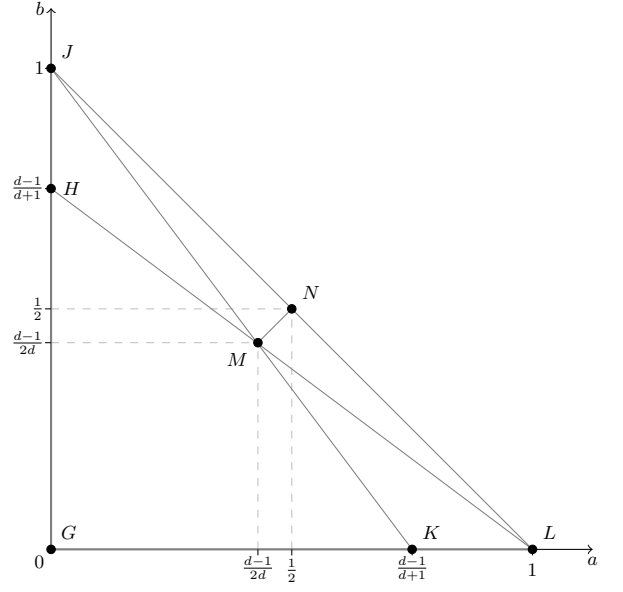

Since $\Tr[\rho_{ab}\rho_{a'b'}]$ is linear in $a'$ and $b'$, and the region $\calN_{c}$ over which $\Tr[\rho_{ab}\rho_{a'b'}]$ is maximized is a polytope, it suffices to only check the vertices of the polytope. These vertices (which we can see in FIG.\ \ref{fig:rhoabNc}) are:
\begin{gather*}
 (0,0),\hspace{3mm} (0,\tfrac{d(c+1)-1}{d}),\hspace{3mm}(\tfrac{1+cd}{d},0),\\
 (\tfrac{1+cd}{d},\tfrac{1-cd}{d}),\hspace{3mm}\text{and}\hspace{3mm}(\tfrac{1-cd}{d},\tfrac{1+cd}{d}).
\end{gather*}
For $\rho_{a,b}$ with $A=\frac{d+1}{d-1}a+b-1$ and $B=\frac{d+1}{d-1}b+a-1$, the trace overlap is given by
\[
 \Tr[\rho_{ab}\rho_{a'b'}] = \frac{1}{d}\left(a'A+b'B + 1-a-b\right).
\]
To maximize this value for a given state $\rho_{ab}$, we split our analysis into the following cases:
\begin{itemize}
 \item $A\leq0$ and $B\leq 0$: The state $\rho_{a'b'}$ that maximizes $h_{\calN_c}(\rho_{ab})$ should have $a'=b'=0$, and thus $h_{\calN_c}(\rho_{ab})=\frac{1-a-b}{d}$. 
 \item $A>0$ and $B\leq 0$: The state $\rho_{a'b'}$ that maximizes $h_{\calN_c}(\rho_{ab})$ should have $b'=0$. Hence the optimal $a'$ is $a'=\tfrac{1+cd}{2}$, and thus $h_{\calN_c}(\rho_{ab})=\frac{cd(d+1)-d+3}{2d(d-1)}a+\frac{cd-1}{2d}b+\frac{1-cd}{2d}$. 
 \item $B>0$ and $A\leq 0$: The state $\rho_{a'b'}$ that maximizes $h_{\calN_c}(\rho_{ab})$ should have $a'=0$. Hence the optimal $b'$ is $b'=\tfrac{d(c+1)-1}{d}$, and thus $h_{\calN_c}(\rho_{ab})=\frac{cd-1}{d^2}a+\frac{cd(d+1)+d-1}{d^2(d-1)}b+\frac{1-cd}{d^2}$.
 \item $0< A\leq B$: The state that maximizes $h_{\calN_c}(\rho_{ab})$ should have $0<a'\leq b'$. Hence the optimal $(a',b')$ is $(\frac{1-cd}{2},\frac{1+cd}{2})$, and thus $h_{\calN_c}(\rho_{ab})=\frac{1-cd}{d(d-1)}a+\frac{1+cd}{d(d-1)}b$.
  \item $0< B\leq A$: The state that maximizes $h_{\calN_c}(\rho_{ab})$ should have $0<b'\leq a'$. Hence the optimal $(a',b')$ is $(\frac{1+cd}{2},\frac{1-cd}{2})$, and thus $h_{\calN_c}(\rho_{ab})=\frac{1+cd}{d(d-1)}a+\frac{1-cd}{d(d-1)}b$.
\end{itemize}

We can now examine the value of the witnesses $W_{\rho_{ab}}$. Since the value of the witness $W_{\rho_{ab}}$ is linear in $a$ and $b$ within each of the regions in FIG.\ \ref{fig:rhoabABregions}, it suffices to check only the extremal points of each of these regions, which we have labeled in FIG.\ \ref{fig:rhoabABpoints}. The values of the witness evaluated at each of these points are listed in Table~\ref{tab:witnessvals}. Unfortunately, the only case when any of these witnesses is an improvement over the negativity is when $\rho_{a,b}$ is the most entangled Werner state, which is just the value of the witness $W'_\text{wer}$ analyzed in the main body of this paper. Hence, the witness $W_\text{Gwer}$ generated by minimizing $W_\tau$ over all generalized Werner states is no better than the witness $W_\text{wer}$ generated by minimizing over the Werner states.

\begin{table}[h!]
 \begin{tabular}{|c|l|l|}
 \hline
 Point & $(a,b)$ & $W_{\rho_{ab}}(\rho,\sigma)$\\
 \hline
 G & $(0,0)$                & $\frac{1}{d}\left(N(\rho)-\Tr[P\sigma^\Gammaminus]\right)$\\
 H & $(0,\tfrac{d-1}{d+1})$ & $\frac{2}{d(d+1)}\left(\frac{d(d+1)-2}{2}N(\rho)-\Tr[(P+P_+)\sigma^\Gammaminus]\right)$\\
 J & $(0,1)$                & $\frac{2}{d(d-1)}\left(\frac{d(c+1)-1}{d}N(\rho) - \Tr[P_+\sigma^\Gammaminus]\right)$\\
 K & $(\tfrac{d-1}{d+1},0)$ & $\frac{2}{d(d+1)}\left(\frac{d(d+1)-2}{2}N(\rho)-\Tr[(P+P_-)\sigma^\Gammaminus]\right)$\\
 L & $(1,0)$                & $\frac{2}{d(d-1)}\left(\frac{1+cd}{2}N(\rho) - \Tr[P_-\sigma^\Gammaminus]\right)$\\
 M & $(\tfrac{d-1}{2d},\tfrac{d-1}{2d})$& $\frac{1}{d^2}\left(N(\rho)-N(\sigma)\right)$\\
 N & $(\tfrac{1}{2},\tfrac{1}{2})$& $\frac{1}{d(d-1)}\left(N(\rho)-\Tr[(P_++P_-)\sigma^\Gammaminus]\right)$\\
 \hline 
 \end{tabular}
\caption{Value of the witness $W_{\rho_ab}(\rho,\sigma)=h_{\calN_c}(\rho_{ab})N(\rho)-\Tr[\rho_{ab}\sigma^\Gammaminus]$ at the points G-N. Here, $c=\min\{\frac{1}{d},\frac{\Tr[\rho^{\Gammaminus\Gammaminus}]}{\Tr[\rho^\Gammaminus]}\}$.}
\label{tab:witnessvals}
\end{table}


\input{witnesspaper6.bbl}

\end{document}

%% file: witnesspaper6.bbl
%